\documentclass[%
twocolumn,
 amsmath,amssymb,
 aps,
floatfix,
]{revtex4-1}

\usepackage{graphicx}
\usepackage{dcolumn}
\usepackage{bm}
\usepackage[utf8]{inputenc}
\usepackage[T1]{fontenc}
\usepackage{romannum}
\usepackage{xr}
\usepackage{natbib}
\usepackage{hyperref}
\usepackage{color}
\usepackage{amsmath}
\usepackage[color=green!40]{todonotes}
\usepackage{soul}
\usepackage{caption}
\usepackage{subfigure}
\usepackage{lipsum}
\setstcolor{red}
\newcommand{\rev}[1]{\textcolor{black}{#1}}
\newcommand{\blue}[1]{\textcolor{black}{#1}}

\newcommand{\Tref}{T_{\rm{L}}}
\newcommand{\ci}{{\bm{c}_i}}
\newcommand{\vi}{{\bm{v}_i}}
\newcommand{\x}{\bm{x}}
\newcommand{\U}{\bm{u}}
\newcommand{\eq}{\rm{eq}}
\newcommand{\dt}{\partial_t^{(1)}}
\newcommand{\dtt}{\partial_t^{(2)}}
\newcommand{\dalpha}{\partial_\alpha^{(1)}}
\newcommand{\dbeta}{\partial_\beta^{(1)}}
\newcommand{\dgamma}{\partial_\gamma^{(1)}}
\newcommand{\ab}{\alpha\beta}
\newcommand{\abg}{\alpha\beta\gamma}
\newboolean{bib}
\setboolean{bib}{false}
\newcommand{\ifindex}{\ifthenelse{\boolean{bib}}}
\begin{document}
\title{Multi-scale analysis of the particles on demand kinetic \rev{model}}

\author{Ehsan Reyhanian}
\email{ehsanr@ethz.ch}
\affiliation{Department of Mechanical and Process Engineering, ETH Zurich, 8092 Zurich, Switzerland}

\begin{abstract}
We present a thorough investigation of the Particles on Demand (PonD) kinetic \rev{model}.
After a brief introduction of the method, an appropriate multi-scale analysis is carried out
to derive the hydrodynamic limit of the model. In these analysis, the effect of the time-space
dependent co-moving reference frames are taken into account. This could be regarded as a
generalization of conventional Chapman-Enskog analysis applied to the Lattice Boltzmann (LB)
models which feature global constant reference frames. Further simulations of target benchmarks
provide numerical evidence confirming the theoretical predictions.

\end{abstract}

\maketitle
\section{Introduction}

The lattice Boltzmann method (LBM) has developed as an essential tool in computational fluid 
dynamics \cite{succi2018lattice,kruger2017lattice}. The ability of this method in various applications
such as multiphase \cite{Sbragaglia2006,Biferale2012,Benzi2009,MazloomiM2015}, micro \cite{Kunert2007,Hyvaluoma2008} 
and turbulent flows \cite{Atif2017,dorschner2016entropic} has long been proven, attracting many researchers to 
extend the merits of this kinetic-based method. As one of the most fundamental fields in fluid dynamics, 
compressible flows have been the focus of significant research efforts, leading to development of "gas dynamics".
The compressibility of a gas and the thermodynamics of such mediums allow shock waves and discontinuous solutions, which
require special treatments in numerical studies \cite{pirozzoli2011numerical}. \\

While the LBM has been extensively used in the incompressible flow regime \cite{succi2018lattice}, its application
in compressible flows is still an open field of study, directing researchers towards developing various 
models \cite{Wilde2020,Feng2019,Frapolli2015,Prasianakis2008,Frapolli2016,frapolli2016entropic}.
Considering that the restrictions in the conventional LB models are mainly due to the fixed velocity sets \cite{succi2018lattice},
the idea of shifted lattices was first introduced in \cite{Frapolli2016}, which found to be
significantly useful in increasing the range of performance of
LB simulations of supersonic flows \cite{Frapolli2016}. 
In this method, the peculiar velocities $\ci$ known for each type of lattice \cite{shyam2006prl} are 
shifted by a constant to mitigate the errors associated
with the violation of Galilean-invariance
\begin{align}
\vi = \ci + \bm{U}.
\end{align}
While the model is able to operate well in uni-lateral flows such as a shock-tube, it can not be used 
in  general setups, where a wide range of temperatures and velocities might emerge.
To overcome this, the idea of projecting particles to the co-moving reference frame led to
the development of the Particles on Demand kinetic \rev{model} \cite{pond}. In the so-called PonD method, 
The peculiar velocities are regarded as the relative velocities with respect to the co-moving reference
frame $\lambda=\{\bm{u},T\}$, where 
$\bm{u}$ is the local velocity of the flow and $T$ is the local temperature. The new definition
of the discrete velocities revokes the 
known restrictions on the range of velocity and temperature in LBM applications. This opens a
novel perspective into the world of computational kinetic methods,
especially for simulation of compressible flows. However, in PonD, the range of complexity rises as
well as its ability to span a wide range of applications, which the former 
models were insufficient or computationally non-efficient to provide accurate solutions.
For example, with the new realization of the discrete velocities in PonD, the advection becomes non-exact
requiring interpolation techniques, where the accuracy and stability of the model will depend 
on the choice of interpolation kernels. This is in contrast to LBM, where a simple and exact point-to-point
streaming step is adopted. Therefore, we will carry out detailed analysis of the model to examine
its range of applicability.\\

In PonD, the discrete velocities are defined as
\begin{align}
\vi = \sqrt{\theta}\ci + \U,
\label{eq:pond/discrete-vel}
\end{align}
where $\theta=T/\Tref$ for an ideal gas. Equation~\eqref{eq:pond/discrete-vel} describes that the
peculiar velocities $\ci$ are first scaled by some definite factor of the square root of the local
temperature and then shifted by the local velocity of the flow. While the former revokes the
restriction on the lattice temperature $\Tref$, the latter results in Galilean-invariance.
The populations corresponding to the reference frame $\lambda=\{T,\U\}$ are denoted by $f_i^\lambda$.

\section{Exact equilibrium}
\label{sec:pond/equilibrium}
To derive the discrete form of the equilibrium, we follow \cite{He-Lou-1997} and consider the non-discrete
velocities $\bm{v} = \sqrt{\theta}\bm{c} + \bm{u}$ in 
the co-moving reference frame $\lambda=\{T,\U\}$. Upon substitution in the Maxwell-Boltzmann
equilibrium function and choosing $\theta = T/\Tref$, one gets
\begin{align}
f^{\lambda,\eq}(\x,\bm{c}) &= \frac{\rho}{(2\pi RT)^{D/2}} \exp \left(  -\frac{(\bm{v}-\bm{u})^2}{2RT}  \right) \nonumber \\
                   &= \frac{\rho}{(2\pi RT)^{D/2}} \exp \left(  -\frac{c^2}{2R\Tref}  \right),
\end{align}
where $D$ stands for the dimension.
We define the phase-space integral
\begin{align}
I = \int \exp\left(-\frac{c^2}{2R\Tref}\right)\Psi(\bm{v})d\bm{v},
\label{eq:pond/integral}
\end{align}
where $\Psi$ is a polynomial in $\bm{v}$. The above integral can be represented by the following
series using the Gaussian-type quadrature
\begin{align}
I = \sum_\alpha W_\alpha \exp\left(-\frac{c_\alpha^2}{2R\Tref}\right)\Psi(\bm{v}_\alpha),
\label{eq:pond/Guassian}
\end{align}
where \rev{$W_{\alpha}$} are the corresponding weights in each direction \rev{${\alpha}$}. To reduce the computations,
it is of interest to 
consider a one-dimensional case. By using the general definition $\Psi(v) = v^m$, where m
is an integer, the integral \eqref{eq:pond/integral} becomes
\begin{align}
I = \sqrt{\theta}\int \exp\left(-\frac{c^2}{2R\Tref}\right)(\sqrt{\theta} c+u)^m  dc.
\end{align}
Introducing the scaling factor $\sqrt{2R\Tref}$ to non-dimensionalize the velocity terms, one can rewrite the latter as
\begin{align}
I = \sqrt{2R\Tref\theta}^{(m+1)}  I_{m,{\hat{u}}/{\sqrt{\theta}}},
\label{eq:pond/I}
\end{align}
where 
\begin{align}
I_{m,a} = \int_{-\infty}^{\infty} \exp\left(-\hat{c}^2\right)( \hat{c}+a)^m  d\hat{c},
\label{eq:pond/phase-space-int}
\end{align}
and the superscript denotes the dimensionless quantities. It is well-known that the following definite 
integral can be expressed in terms of a third-order Hermite formula
\begin{align}
I_{m} = \int_{-\infty}^{\infty} \exp\left(-x^2\right)x^m dx = \sum_{j=1}^{3} \tilde{w}_j x_j^m,
\end{align}
where $x$ is a dummy variable and $x_1 = -\sqrt{3/2}, x_2 = 0, x_3 = \sqrt{3/2}$ are the abscisas
with the corresponding weights of $\tilde{w}_1 = \sqrt{\pi}/6, \tilde{w}_2 = 2\sqrt{\pi}/3, \tilde{w}_3 = \sqrt{\pi}/6$.
 Using the Newton formula, one can expand Eq. \eqref{eq:pond/phase-space-int} as

\begin{align}
I_{m,a} &= \sum_{k=0}^{m} \int_{-\infty}^{\infty} \exp\left(-x^2\right) x^m dx \left( \begin{array}{c} m \\ k \end{array} \right) a^{m-k} \nonumber\\
&= \sum_{k=0}^{m} \sum_{j=1}^3 \tilde{w}_j x_j^k \left( \begin{array}{c} m \\ k \end{array} \right) a^{m-k} \nonumber\\&= \sum_{j=1}^{3} \tilde{w}_j \sum_{k=0}^m  x_j^k \left( \begin{array}{c} m \\ k \end{array} \right) a^{m-k} \nonumber\\
&= \sum_{j=1}^3 \tilde{w}_j \left(x_j+a\right)^m.
\end{align}
Therefore, the peculiar discrete velocities are derived as before
\begin{align}
&c_1 = \sqrt{2R\Tref} \hat{c}_1 = -\sqrt{3R\Tref}, \nonumber\\
&c_2 = 0, \nonumber\\
&c_3 = \sqrt{2R\Tref} \hat{c}_3 = \sqrt{3R\Tref},
\end{align}
which constructs a $D1Q3$ lattice $\{-C,0,C\}$ with the lattice temperature $\Tref = C^2/3$.\\
Finally, Eq. \eqref{eq:pond/I} reduces to
\begin{align}
I &= \sqrt{2RT}\sqrt{2R\Tref}^m \sum_{i=0}^3 \tilde{w}_i \left( \sqrt{\theta} \hat{c}_i + \hat{u} \right)^m, \nonumber\\
  &= \sqrt{2\pi RT}\sum_{i=0}^3 w_i \left( \sqrt{\theta} c_i + u \right)^m, \nonumber\\
  &= \sqrt{2\pi RT}\sum_{i=0}^3 w_i \Psi(v_i),
\end{align}
where $w_i = \tilde{w}_i/\sqrt{\pi}$. Due to the splitting property of the phase-space integral,
i.e. $d\bm{c} = dc_1...dc_D$ the latter formula 
in $D$ dimensions becomes
\begin{align}
  I= \sqrt{2\pi RT}^D\sum_{\alpha=0}^\mathcal{Q} w_\alpha \Psi(v_\alpha),
\end{align}
where $w_\alpha=w_{i,j,...,D}=w_i w_j ...w_D$ is the tensor product of one-dimensional weights, $v_\alpha=v_{(i,j,...,D)}$ and 
$\mathcal{Q} = ij...D$ is the total number of discrete velocities.
Considering Eq. \eqref{eq:pond/Guassian}, the Gaussian weights are obtained as
\begin{align}
  W_\alpha= \sqrt{2\pi RT}^D w_\alpha \exp \left(\frac{c_\alpha^2}{2R\Tref}\right).
\end{align}
Finally, the discretized equilibrium populations are derived from the continuous function as
\begin{align}
  f_\alpha^{eq}&= W_\alpha f^{eq}(\x,c_\alpha) = \sqrt{2\pi RT}^D w_\alpha \exp \left(\frac{c_\alpha^2}{2R\Tref}\right)\nonumber\\
    &= \sqrt{2\pi RT}^D w_\alpha \exp \left(\frac{c_\alpha^2}{2R\Tref}\right) \frac{\rho}{(2\pi RT)^{D/2}} \exp \left(  -\frac{c_\alpha^2}{2R\Tref}  \right)\nonumber,\\
 &= \rho w_\alpha,
\end{align}
 which is exact and free of velocity terms and hence, the Galilean invariance is ensured.

\rev{\section{Analysis of Particles on Demand}}
\label{sec:pond/analysis}
In this section, we analyze the kinetic equations in the PonD framework. 
After a brief introduction of the method, we demonstrate how to derive the recovered thermo-hydrodynamic limit. 
Namely, we conduct the Chapman-Enskog analysis by expanding the kinetic equations into multiple levels of time and space scales. 
Finally, we derive the recovered range of the Prandtl number and present an order verification study.\\

\subsection{Kinetic equations}
\label{sec:pond/kinetic-eq}
Similar to LBM, the kinetic equations can split into two main parts; Collision \rev{using 
the Bhatnagar-Gross-Krook (BGK) model}, with exact equilibrium-populations
\begin{align}
f_i^*(\x,t) = f_i(\x,t) + \omega(\rho w_i - f_i)_{(\x,t)},
\end{align} 
where  $f_i^*(\x,t)$ is the post-collision populations, which are computed at the gauge $\lambda=\lambda(\x,t)$
\rev{ and $\omega$ is the relaxation parameter}. 
Next, the streaming step is conducted via the semi-Lagrangian method, where the 
information at the monitoring point $(\x,t)$ is updated by traveling back through the 
characteristics to reach the departure point $\x_{id} = \x - \vi\delta t$. However, due to the dependency of 
the discrete velocities \eqref{eq:pond/discrete-vel} on the local flow field, the departure point may be located
off the grid points. This is in contrast to LBM, where the lattice provides exact streaming along the links. Hence,
the information at the departure point must be interpolated through the collocation points. Furthermore, 
in order to be consistent, the populations at the departure point must be in the same reference frame as the
monitoring point. Hence, the populations at the collocation points are
first transformed to the gauge of the monitoring point and then are interpolated \cite{pond}.
Finally, the advection step is indicated by
\begin{align}
f_i(\x,t) = \sum_{p=0}^{N-1} \Lambda(\x_d-\x_p)\mathcal{G}_{\lambda_p}^{\lambda}f^{*\lambda_P}(\x_p,t),
\label{eq:SL-advection}
\end{align}
where $\bm{x}_p$, $p=0,...,N-1$ denote the collocation points (grid points) and $\Lambda$ is the interpolation kernel.
As mentioned before, the populations are transformed using the transformation Matrix $\mathcal{G}$.
In general, a~set of populations at gauge $\lambda$ can be transformed to another gauge $\lambda^\prime$ by matching $Q$ linearly independent moments:
\begin{align}
    \bm{M}_{mn}^\lambda = \sum_{i=1}^Q f_i^\lambda v_{ix}^m v_{iy}^n,
\label{eq:independent-moments}
\end{align}
where $m$ and $n$ are integers. This may be written in the matrix product form as $\bm{M}^\lambda = \mathcal{M}_\lambda f^\lambda$ where $\mathcal{M}$ is the $Q\times Q$ linear map. 
Requiring that the moments must be independent from the choice of the reference frame, leads to the matching condition:
\begin{align}
    \mathcal{M}_{\lambda^\prime}f^{\lambda^\prime} = \mathcal{M}_{\lambda}f^{\lambda}, 
\label{eq:mom-invariant}
\end{align}
which {yields} the transformed populations:
\begin{align}
    f^{\lambda^\prime} = \mathcal{G}_\lambda^{\lambda^\prime}f^\lambda=\mathcal{M}_{\lambda^\prime}^{-1}\mathcal{M}_{\lambda}f^{\lambda}.
\end{align}
\\

Finally, the macroscopic values are evaluated by taking the pertinent moments
\begin{align}
\rho &= \sum_i f_i,\label{eq:macro-moments-rho}\\
\rho\U &= \sum_i f_i\vi,\label{eq:macro-moments-vel}\\
\rho u^2 + D\rho T &= \sum_i f_i v_i^2.\label{eq:macro-moments-temp}
\end{align}
The implicitness in the above equations require a predictor-corrector step to 
find the co-moving reference frame. Hence, the advection step is repeated by
imposing the new evaluated velocity and temperature until the convergence is 
achieved.
\rev{To this end, discrete velocities \eqref{eq:pond/discrete-vel} at each monitoring point $(\bm{x},t)$ are initially 
set relative to the gauge $\lambda_0=\{T_0,\bm{u}_0\}$, where 
$\bm{u}_0=\bm{u}(\bm{x},t-\delta t)$ and $T_0=T(\bm{x},t-\delta t)$ are known from the 
previous time step. Constructing the initial discrete velocities $v_i^0=\sqrt{\theta_0}\ci+\U_0$, 
the advection \eqref{eq:SL-advection} is followed to compute the populations $f_i^{\lambda_0}(\x)$.
Using Eqs. \eqref{eq:macro-moments-rho}-\eqref{eq:macro-moments-temp}, the 
new macroscopic quantities are evaluated to define the corrected gauge $\lambda_1=\{T_1,\U_1\}$,
which results in the corrected velocities $v_i^1$ and consequent populations $f_i^{\lambda_1}$.
The iterations will continue until the reference frame
is converged to a fixed value $\lambda_{\infty}={\rm lim}_{n\to\infty} \{T_n,\U_n\}$.
In the limit of the co-moving reference frame, the computed velocity $\U_{\infty}=\U(\x,t)$ and temperature 
$T_{\infty}=T(\x,t)$ by moments \eqref{eq:macro-moments-vel} and \eqref{eq:macro-moments-temp}
are equal to those defined as the reference 
frame $\lambda_{\infty}=\lambda(\x,t) = \{T(\x,t),\U(\x,t)\}$, i.e.}
\begin{align}
\sum_i f_i^{\lambda}\U &= \sum_i f_i^{\lambda}(\sqrt{T/\Tref}\ci+\U),\label{eq:comoving-1}\\
\sum_i f_i^{\lambda}(u^2 + D T) &= \sum_i f_i^{\lambda} ||\sqrt{T/\Tref}\ci+\U||^2.
\label{eq:comoving-2}
\end{align}
For more details, see \cite{pond}. 
\rev{A convergence analysis for the iterative algorithm of "predictor-corrector" is provided in the appendix.}

\subsection{Chapman-Enskog analysis}
\label{sec:pond/ce-analysis}
\begin{figure}
\centering
\includegraphics[clip, trim = 2cm 1cm 2cm 0.5cm, width=\linewidth]{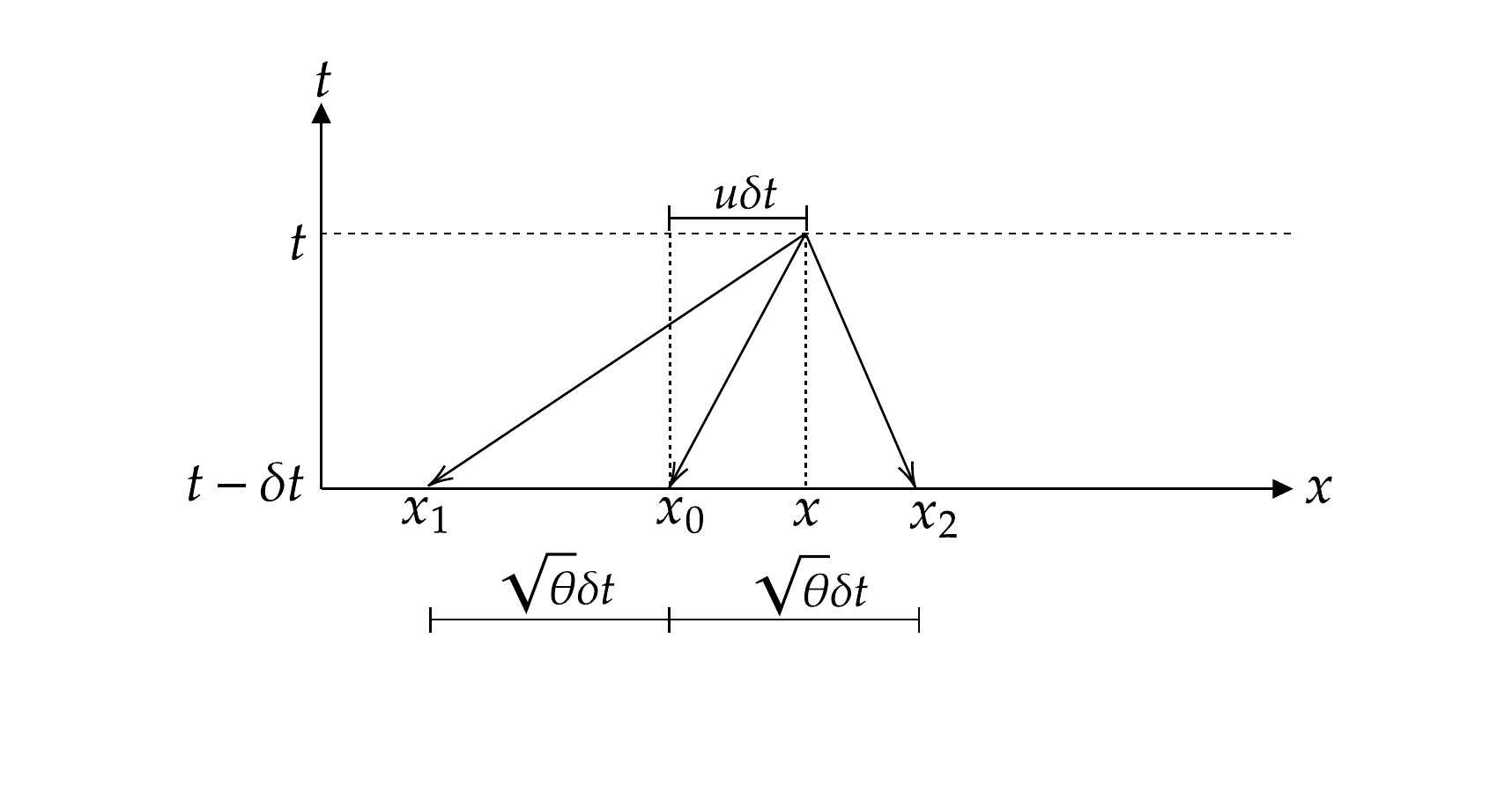}
\caption{Schematic of the semi-Lagrangian advection and the location of the departure points.}
\label{fig:streaming}
\end{figure}

In this section, we aim at recovering the hydrodynamic limit of the model. 
\rev{Before we begin, it is important to note that due to time-space dependent discrete velocities, the non-commutativity relation
$v_{i\alpha}\partial_{\alpha}f_i\neq \partial_{\alpha}(v_{i\alpha}f_i)$ is taken into account at each step of the following analysis.}\\
We assume that the co-moving reference frame $\lambda(\x,t)$ has been reached at the
monitoring point. In other words, Eqs. \eqref{eq:comoving-1} and \eqref{eq:comoving-2} are legit.
For simplicity, we first neglect the interpolation process and recast the advection equation as 
\begin{align}
f_i^{\lambda}(\x,t) = \mathcal{G}_{\lambda_i}^\lambda f_i^*(\x_{id},t-\delta t),
\label{eq:streaming}
\end{align} 
where $\lambda_i = \lambda(\x_{id},t-\delta t)$ is the corresponding co-moving reference frame 
at each departure point. Figure \ref{fig:streaming} 
illustrates the semi-Lagrangian advection and the departure points using the $D1Q3$ lattice. 
By definition, Eq. ~\eqref{eq:streaming} is recast 
into the following form
\begin{align}
f_i^{\lambda}(\x,t) = \mathcal{M}_{i,\lambda}^{-1} M^*(\x_{id},t-\delta t),
\label{eq:streaming-recast}
\end{align} 
where the dummy indices are dropped in the right hand side and
\begin{align}
M^* = \mathcal{M}f^* = M + \omega(M^{\eq} - M),
\end{align}
is the post-collision moments. Note that since the equilibrium populations are exact,
the equilibrium moments $M^{\eq}$ coincide with the Maxwell-Boltzmann moments.

In the following, we also drop the superscript $\lambda$ for simplicity.
Using the Taylor expansion up to third-order one can write 
\begin{align}
&M^*(\x-\vi\delta t,t-\delta t) = M^*(\x,t) - \delta t D_i M^* \nonumber\\
&+ \frac{\delta t^2}{2}(D^2_i M^* - D_i v_{i\alpha}\partial_{\alpha}M^*)+\mathcal{O}(\delta t^3),
\label{eq:taylor}
\end{align} 
where $D_i = \partial_t + v_{i\alpha} \partial_\alpha$ is the material derivative.
Finally, substituting the expansion \eqref{eq:taylor} into Eq. ~\eqref{eq:streaming-recast} results in
\begin{align}
&\delta t D_i f_i - \frac{\delta t^2}{2}D^2_i f_i = -\omega f_i^{\rm{neq}} + \delta t D_i(\omega f_i^{\rm{neq}}) \nonumber\\
& + \delta t D_i\mathcal{M}_i^{-1}M^* 
- \frac{\delta t^2}{2}D^2_i(\omega f_i^{\rm neq}) - \frac{\delta t^2}{2}\mathcal{M}^{-1}_iD_i v_{i\alpha}\partial_{\alpha} M^* \nonumber\\
&- \frac{\delta t^2}{2}D_i\left(D_i \mathcal{M}_i^{-1}M^*\right) - \frac{\delta t^2}{2}D_i\mathcal{M}_i^{-1}D_i M^*.
\label{eq:evolution-f-1}
\end{align} 
By applying the operator $D_i$ upon the latter equation and neglecting the higher order terms $\mathcal{O}(\delta t^3)$, we get
\begin{align}
\frac{\delta t^2}{2}D^2_i f_i &= - \frac{\delta t}{2} D_i(\omega f_i^{\rm{neq}}) 
+ \frac{\delta t^2}{2}D^2_i(\omega f_i^{\rm neq}) \nonumber\\
&+ \frac{\delta t^2}{2}D_i\left(D_i \mathcal{M}_i^{-1}M^*\right).
\label{eq:evolution-f-2}
\end{align} 
Eventually, substituting Eq. \eqref{eq:evolution-f-2} from Eq. \eqref{eq:evolution-f-1} yields to
\begin{align}
D_i f_i= -\frac{\omega}{\delta t}f_i^{\rm{neq}} +  D_i\left(\frac{\omega}{2}f_i^{\rm{neq}}\right) + D_i\mathcal{M}_i^{-1}M^* \nonumber\\
 - \frac{\delta t}{2}D_i\mathcal{M}_i^{-1}D_i M^* - \frac{\delta t}{2}\mathcal{M}_i^{-1}D_i v_{i\alpha}\partial_{\alpha} M^*.
\end{align} 
To start the analysis, first the following expansions are introduced
\begin{align}
f_i &= f_i^{(0)} + \epsilon f_i^{(1)} + \epsilon^2 f_i^{(2)},\nonumber\\
M^* &= M^{*(0)} + \epsilon M^{*(1)} + \epsilon^2 M^{*(2)},\nonumber\\
\partial_t &=\epsilon \partial_t^{(1)} + \epsilon^2 \partial_t^{(2)},\nonumber\\
\partial_\alpha &=\epsilon \partial_\alpha^{(1)}.
\label{eq:expansions}
\end{align}
Rearranging the equations and collecting the corresponding terms on each order yields to
\begin{align}
\mathcal{O}(\epsilon^0):\hspace{1mm}
f_i^{(0)} = f_i^{\eq} \to M^{(0)} = M^{\eq},
\end{align}
\begin{align}
\mathcal{O}(\epsilon^1):\hspace{1mm}
&D_i^{(1)} f_i^{(0)}- D_i^{(1)}\mathcal{M}_i^{-1}M^{(0)}= -\frac{\omega}{\delta t}f_i^{(1)}
\label{eq:first-order},
\end{align}
\begin{align}
\mathcal{O}(\epsilon^2):\hspace{1mm}
&\dtt f_i^{(0)} + D_i^{(1)}\left[\left(1-\frac{\omega}{2}\right)f_i^{(1)}\right] = -\frac{\omega}{\delta t}f_i^{(2)} \nonumber\\
&+ \dtt\mathcal{M}_i^{-1}M^{(0)}
 + D_i^{(1)} \mathcal{M}_i^{-1}\left(1-\frac{\omega}{2}\right)M^{(1)}\nonumber\\
  &- \frac{\delta t}{2}\mathcal{M}_i^{-1}D_i^{(1)}\left(\mathcal{M} V_{\alpha}\mathcal{M}^{-1}\right)\partial_{\alpha}M^{(0)},
\label{eq:second-order}
\end{align}
where $V_{\alpha}={\rm diag}(v_{\alpha})$.
Reminding $f_i^{(0)} = \mathcal{M}_i^{-1}M^{(0)}$, Eq. \eqref{eq:first-order} is rewritten as
\begin{align}
\mathcal{M}_i^{-1}\dt M^{(0)} + v_{i\alpha}\mathcal{M}_i^{-1}\dalpha M^{(0)} = -\frac{\omega}{\delta t}f_i^{(1)},
\label{eq:first-order-modified}
\end{align}
where we can derive the first order evolution equation for the equilibrium moments by multiplying both sides of
Eq.~\eqref{eq:first-order-modified} by $\mathcal{M}$ and reminding $M^{(1)}=\mathcal{M}f^{(1)}$
\begin{align}
\mathcal{D}^{(1)} M^{(0)} = -\frac{\omega}{\delta t}M^{(1)},
\label{eq:first-order-final}
\end{align}
where $\mathcal{D} = \partial_t + \mathcal{M}V_{\alpha}\mathcal{M}^{-1}\partial_{\alpha}$.
Finally, the second-order kinetic equation \eqref{eq:second-order} can be rearranged to
\begin{align}
\dtt M^{(0)} &+ \mathcal{D}^{(1)}\left[(1-\frac{\omega}{2})M^{(1)}\right] = -\frac{\omega}{\delta t}M^{(2)} \nonumber\\
&-\frac{\delta t}{2}\mathcal{D}^{(1)}\left(\mathcal{M}V_{\alpha}\mathcal{M}^{-1}\right)\partial_{\alpha}M^{(0)},
\label{eq:second-order-final}
\end{align}
where $M^{(2)}=\mathcal{M}f^{(2)}$.


\subsection{Conservation equations}
With the split kinetic equations at three different orders, we are now able \rev{to} derive the hydrodynamic limit of the present kinetic model.
However, due to the dependence of the multi-scale kinetic equations on the linear mapping matrix $\mathcal{M}$ and its corresponding inversion,
we shall specify a lattice to proceed with the analysis. In the following, 
we consider the most commonly used lattices in one and two-dimensional applications, i.e.
$D1Q3, D1Q5, D2Q9$ and $D2Q25$, where 
the peculiar velocities $\ci$  and the 
lattice reference temperature $\Tref$ in \eqref{eq:pond/discrete-vel} are known for each of them \cite{shyam2006prl}.

\subsubsection{$D1Q3$}
The three linearly independent moments in \eqref{eq:independent-moments} are
\begin{align}
M_{00} &= \sum_i f_i,\nonumber\\
M_{10} &= \sum_i f_i v_i,\nonumber\\
M_{20} &= \sum_i f_i v_i^2,
\end{align}
which all are conserved moments and coincide with their counterpart equilibrium ones.
Hence, the mass, momentum and total energy conservation implies that $M^{(1)}=M^{(2)}=[0,0,0]^\dagger$.
The inversion of the mapping matrix is obtained as
\begin{align}
\mathcal{M}^{-1} = \left[
\begin{tabular}{ccc}
$1-\frac{u^2}{\theta}$ & $\frac{2u}{\theta}$ & $\frac{-1}{\theta}$\\
$\frac{u^2-\sqrt{\theta}u}{2\theta}$ &	$\frac{\sqrt{\theta}-2u}{2\theta}$ & $\frac{1}{2\theta}$\\
$\frac{u^2+\sqrt{\theta}u}{2\theta}$ &	$\frac{-\sqrt{\theta}-2u}{2\theta}$ & $\frac{1}{2\theta}$
\end{tabular}
\right],
\label{eq:mapping-inverse}
\end{align}
where it is observed that
\begin{align}
&\sum_{i} \mathcal{M}^{-1}_{ij} = \Bigg\{
\begin{tabular}{ll}
1, & $j=1$, \\
0, & {\rm otherwise}.
\label{eq:inverse-moment-prop}
\end{tabular}
\end{align}
Finally, the first order equations are recovered from Eq. \eqref{eq:first-order-final} as
\begin{align}
\dt\left[
\begin{tabular}{c}
$\rho$ \\ $\rho u$ \\ $\rho u^2 + \rho\theta\Tref$
\end{tabular}
\right]
+ \dalpha\left[
\begin{tabular}{c}
$\rho u$ \\ $\rho u^2 + \rho\theta\Tref$ \\ $2\rho u H$
\end{tabular}
\right]
= \left[
\begin{tabular}{c}
0\\0\\0
\end{tabular}
\right],
\label{eq:d1q3-first-order}
\end{align}
where $H=\theta/2 + u^2/2$ is the total enthalpy and $h=\theta/2$ is the specific enthalpy, which implies
$C_p=3/2$ for an ideal gas.
Similarly, the second order equations are obtained from Eq. \eqref{eq:second-order-final} 
\begin{align}
\dtt\left[
\begin{tabular}{c}
$\rho$ \\ $\rho u$ \\ $\rho u^2 + \rho\theta\Tref$
\end{tabular}
\right]
= -\frac{\delta t}{2}\left[ 
\begin{tabular}{c}
0\\$X_M^{(1),(1)}$\\$X_E^{(1),(1)}$
\end{tabular}
\right],
\label{eq:d1q3-second-order}
\end{align}
where 
\begin{align}
X_M^{(1),(1)} &= \rho \dalpha u \dalpha\theta, \\
X_E^{(1),(1)} &= \rho\dt\theta\dalpha u + \rho\dalpha \theta(\dt u+3\dalpha u^2),
\end{align}
and the double superscript denotes the product of two first-order terms.
Finally, the hydrodynamic equations are recovered by collecting the first and second order 
equations \eqref{eq:d1q3-first-order} and \eqref{eq:d1q3-second-order} and reminding the expansions \eqref{eq:expansions}

\begin{align}
\partial_t\left[
\begin{tabular}{c}
$\rho$ \\ $\rho u$ \\ $\rho u^2 + \rho\theta\Tref$
\end{tabular}
\right]
+ \partial_{\alpha}\left[
\begin{tabular}{c}
$\rho u$ \\ $\rho u^2 + \rho\theta\Tref$ \\ $2\rho u H$
\end{tabular}
\right]
= \left[
\begin{tabular}{c}
0\\$-X_M$\\$-X_E$
\end{tabular}
\right],
\end{align}
where
\begin{align}
X_M &= \delta t \rho \partial_{\alpha}u\partial_{\alpha}\theta, \\
X_E &= \frac{\delta t}{2}\left(\partial_{\alpha}(\rho\theta)\partial_{\alpha}u^2-\partial_{\alpha}\theta\partial_{\alpha}(\rho\theta\Tref)\right),
\end{align}
are the error terms in the momentum and energy equations, respectively. 
As a conclusion, the thermo-hydrodynamic equations for the $D1Q3$ lattice are recovered
as  the one-dimensional compressible Euler equations (vanishing viscosity) with error terms of
$\mathcal{O}(\delta t)$ in the momentum and energy equations. 
\subsubsection{$D1Q5$}
In this section, we consider the $D1Q5$ lattice with the discrete velocities
$\mathcal{\bm{C}}=\{0,\pm m, \pm n\}$, where $m=rn$ and $r=(\sqrt{5}-\sqrt{2})/\sqrt{3}$
\rev{is the ratio of the roots of the 
fifth-order Hermite polynomial \cite{shyam2006prl}.}
The weights and the lattice reference temperature are defined as
\begin{align}
w_0       &= \frac{-3r^4-3+54r^2}{75r^2},\\
w_{\pm m} &= \frac{9r^4-6-27r^2}{300r^2(r^2-1)},\\
w_{\pm n} &= \frac{9-6r^4-27r^2}{300(1-r^2)},\\
\Tref     &= \frac{m^2(r^2+1)}{10r^2},
\end{align}
where we choose $m=1$.
The independent system of moments are $M=[M_{00},M_{10},M_{20},M_{30},M_{40}]^{\dagger}$ 
with the non-equilibrium moments 
\rev{
\begin{align}
M^{(k)}=[0,0,0,\sum f_i^{(k)}v_i^3,\sum f_i^{(k)}v_i^4]^{\dagger}; k=1,2.
\end{align}
}
Once again, we observe that the relation \eqref{eq:inverse-moment-prop} holds for
this lattice structure as well.
According to Eq. \eqref{eq:first-order-final}, the first order equations are derived 
correctly as in \eqref{eq:d1q3-first-order}. On the other hand, Eq. \eqref{eq:second-order-final} gives 
the second-order equations as

\begin{align}
\dtt\left[
\begin{tabular}{c}
$\rho$ \\ $\rho u$ \\ $\rho u^2 + \rho\theta\Tref$
\end{tabular}
\right]
+
\dalpha\left[
\begin{tabular}{c}
$0$ \\ $0$ \\ $(1-\omega/2)q^{(1)}$
\end{tabular}
\right]
= \left[ 
\begin{tabular}{c}
0\\0\\0
\end{tabular}
\right],
\label{eq:d1q5-second-order}
\end{align}

where
\begin{align}
q^{(1)} = \sum_i f_i^{(1)}v_i^3 = -\frac{\delta t}{\omega}\left(\dt Q^{\eq} + \dalpha R^{\eq}\right),
\end{align}
is the non-equilibrium heat flux derived from Eq. \eqref{eq:first-order-final}
and 
\begin{align}
Q^{\eq} = \sum_i \rho w_i v_i^3,\\
R^{\eq} = \sum_i \rho w_i v_i^4,\label{eq:Req}
\end{align}
are the equilibrium high-order moments coinciding with their Maxwell-Boltzmann expressions.
It is straightforward to show
\begin{align}
q^{(1)} = -\frac{2\delta t}{\omega}\left((3-\gamma)pu\dalpha u + pC_p\dalpha T \right).
\label{eq:noneq-heat-d1q5}
\end{align}
Since using a single population leads to a fixed specific heat $\gamma=(D+2)/D$,
the viscous part in \eqref{eq:noneq-heat-d1q5} vanishes, while the Fourier heat flux
is retained. This is in contrast to the $D1Q3$ lattice, where due to the same number of 
velocities and conservation laws, the Fourier heat flux in the energy equation 
vanishes as well as the viscous terms in the momentum and energy equations.
Finally, the thermohydrodynamic equations recovered by using the $D1Q5$ lattice are obtained as

\begin{align}
\partial_t\left[
\begin{tabular}{c}
$\rho$ \\ $\rho u$ \\ $\rho u^2 + \rho\theta\Tref$
\end{tabular}
\right]
+ \partial_{\alpha}\left[
\begin{tabular}{c}
$\rho u$ \\ $\rho u^2 + \rho\theta\Tref$ \\ $2\rho u H$
\end{tabular}
\right]
= \left[
\begin{tabular}{c}
0\\$0$\\$2\partial_{\alpha}(k\partial_{\alpha}T)$
\end{tabular}
\right],
\end{align}
where $k=(1/\omega-1/2)p\delta t C_p$ is the conductivity and $C_p=3/2$.

The most distinctive feature of the recovered equations are the absence of 
error terms in the momentum and energy equations.
\rev{ Although the Galilean-invariance of $D1Q5$ models has been verified in isothermal setups \cite{shyam2006prl},
here we observe a somewhat different behavior. The adaptive construction of discrete velocities in PonD
guaranties Galilean-invariance even with the $D1Q3$ lattice. Having the exact equilibrium, all the recovered equilibrium moments
up to fourth-order (Eq. \eqref{eq:Req}), match with their Maxwell-Boltzmann counterparts. However, we observe that the insufficiency of the
mapping matirx $\mathcal{M}$ and its inversion in the $D1Q3$ lattice is responsible for the generated errors 
(see Eq. \eqref{eq:second-order}). According to the invariant-moment rule \eqref{eq:mom-invariant}, 
the sufficiency of linearly independent moments is crucial for a meaningful transformation between two reference frames. 
Not having met this criteria, the $D1Q3$ (and its two dimensional tensor product as we will see later) is unable 
to provide an error-free transformation. On the other hand, due to its sufficient system of moments,
the $D1Q5$ lattice does not introduce errors during the transformation and together with the fully recovered equilibrium moments, 
the hydrodynamic equations are derived in their correct form.}

\subsubsection{$D2Q9$}
The $D2Q9$ lattice can be considered as the tensor product of two $D1Q3$ lattices. 
The independent moment system in this type of lattice structure is
\begin{align}
M=[M_{00},M_{10},M_{01},M_{11},M_{20},M_{02},M_{21},M_{12},M_{22}]^{\dagger},
\end{align}
where the non-equilibrium moments are
\begin{align}
M^{(k)}=[0,0,0,M_{11}^{(k)},M_{20}^{(k)},M_{02}^{(k)},M_{21}^{(k)},M_{12}^{(k)},M_{22}^{(k)}]^{\dagger},
\end{align}
and the  conservation of energy implies $M_{20}^{(k)}+M_{02}^{(k)}=0$.

The first-order equations are derived as

\begin{align}
\dt\left[
\begin{tabular}{c}
$\rho$ \\ $\rho u_{\alpha}$ \\ $P^{\eq}_{\alpha\alpha}$
\end{tabular}
\right]
+ \dbeta\left[
\begin{tabular}{c}
$\rho u_{\beta}$ \\ $P^{\eq}_{\alpha\beta}$ \\ $2\rho u_{\beta} H$
\end{tabular}
\right]
= \left[
\begin{tabular}{c}
0\\0\\0
\end{tabular}
\right],
\label{eq:d2q9-first-order}
\end{align}
where\begin{align}
P^{\eq}_{\ab} = \rho u_{\alpha}u_{\beta} + \rho\theta\Tref\delta_{\ab},
\end{align}
is the equilibrium pressure tensor.
The second-order equations are obtained as

\begin{align}
\dtt\left[
\begin{tabular}{c}
$\rho$ \\ $\rho u_{\alpha}$ \\ $P_{\alpha\alpha}^{\eq}$
\end{tabular}
\right]
+
\dbeta\left[
\begin{tabular}{c}
$0$ \\ $(1-\omega/2)P^{(1)}_{\ab}$ \\ $(1-\omega/2)Q^{(1)}_{\alpha\ab}$
\end{tabular}
\right]
= \left[ 
\begin{tabular}{c}
0 \\ $\mathcal{O}(\delta t)$ \\ $\mathcal{O}(\delta t)+3q$
\end{tabular}
\right],
\label{eq:d2q9-second-order}
\end{align}

where $P_{\ab}^{(1)}$ is the non-equilibrium
pressure tensor derived from Eq. \eqref{eq:first-order-final}
\begin{align}
P_{\ab}^{(1)} = -\left(\frac{\delta t}{\omega}\right)\left(\dt P^{\eq}_{\ab} + \dgamma Q^{\eq}_{\abg} \right),
\label{eq:noneq-pressure}
\end{align}
and 
\begin{align}
Q^{\eq}_{\abg} = \rho u_{\alpha}u_{\beta}u_{\gamma} + \rho\theta\Tref(u_{\alpha}\delta_{\beta\gamma} + u_{\beta}\delta_{\alpha\gamma} + u_{\gamma}\delta_{\alpha\beta}).
\end{align}
Using the first-order equations \eqref{eq:d2q9-first-order}, it can be shown that
\begin{align}
\dt P_{\ab}^{\eq} + \dgamma Q^{\eq}_{\abg} = &p\left(\dalpha u_{\beta} + \dbeta u_{\alpha} \right) \nonumber\\
&+ (p-\rho c_s^2)\dgamma u_{\gamma}\delta_{\ab},
\end{align}
where $c_s^2=\gamma T$ is the speed of sound of an ideal-gas.\\

The second-order equation for the energy part \eqref{eq:d2q9-second-order} is originally derived as
\begin{align}
&\dtt (P_{xx}^{\eq}+P_{yy}^{\eq}) + \partial_x^{(1)}\left[\left(1-\frac{\omega}{2}\right)Q_{xyy}^{(1)}\right] \nonumber\\
&+ \partial_y^{(1)}\left[\left(1-\frac{\omega}{2}\right)Q_{xxy}^{(1)}\right] + 3u\partial_x^{(1)}\left[\left(1-\frac{\omega}{2}\right)P_{xx}^{(1)}\right]\nonumber\\
&+3v\partial_y^{(1)}\left[\left(1-\frac{\omega}{2}\right)P_{yy}^{(1)}\right] = \mathcal{O}(\delta t),
\end{align}
while the closure relation
\begin{align}
Q^{(1)}_{nnn} = 3u_n P^{(1)}_{nn},\ n=x,y
\label{eq:closure}
\end{align}
has been used to render the final equation in a concise form. As a result, the error term $3q$ appears
in the R.H.S of energy equation, where
\begin{align}
q = \left( 1- \frac{\omega}{2}\right) \left[P^{(1)}_{xx}\partial_x^{(1)} u + P^{(1)}_{yy}\partial_y^{(1)} v\right].
\end{align}
However, the non-equilibrium heat flux $Q_{\alpha\alpha\beta}^{(1)}$ is computed in two separate steps.
While the terms $Q_{xxy}^{(1)}$ and $Q_{xyy}^{(1)}$ are included in the non-equilibrium system of moments
in \eqref{eq:first-order-final}, the diagonal elements $Q_{xxx}^{(1)}$ and $Q_{yyy}^{(1)}$ are slaved by the
closure equation \eqref{eq:closure}. Consequently, the final form of the non-equilibrium heat flux is derived as
\begin{align}
Q_{\alpha\alpha\beta}^{(1)} = -\frac{\delta t}{\omega} \left[\dt Q_{\alpha\alpha\beta}^{\eq}
+ \dalpha R_{\ab}^{\eq}-3\rho\theta\Tref\dbeta(\theta\Tref)\right],
\label{eq:noneq-fourier}
\end{align}
where 
\begin{align}
R_{\ab}^{\eq} = 2\rho u_{\alpha}u_{\beta}(H+\theta\Tref)+2\rho\theta\Tref H\delta_{\ab},
\end{align}
is the fourth-order equilibrium moment and from Eq. \eqref{eq:d2q9-first-order} one can compute
\begin{align}
\dt Q_{\alpha\alpha\beta}^{\eq} + \dalpha R_{\ab}^{\eq} &= 2pu_{\alpha}\left(\dalpha u_{\beta} + \dbeta u_{\alpha} \right) \nonumber\\
&+ 2(p-\rho c_s^2)\dgamma u_{\gamma}u_{\beta} + 2p\dbeta h,
\end{align}
where $h=(D/2+1)\theta\Tref$.
In an interesting note, we observe that the insufficiency of the 
diagonal elements of the third-order non-equilibrium moment has caused an anomaly in the appearance of the
non-equilibrium heat flux \eqref{eq:noneq-fourier}. The non-conventional term in the R.H.S of Eq. \eqref{eq:noneq-fourier}
will contribute to the Fourier heat flux and will alter the value of the Prandtl number as we will see later. 
On a separate comment, we note that similar to the $D1Q3$ case, there exist error terms with the order 
of $\mathcal{O}(\delta t)$ in the momentum and energy equations.

Finally, the hydrodynamic equations for the $D2Q9$ lattice are recovered as

\begin{align}
\partial_t\left[
\begin{tabular}{c}
$\rho$ \\ $\rho u_{\alpha}$ \\ $\rho E$
\end{tabular}
\right]
+
\partial_{\beta}\left[
\begin{tabular}{c}
$\rho u_{\beta}$ \\ $\rho u_{\alpha}u_{\beta} + p\delta_{\ab}+\tau_{\ab}$ \\ $\rho u_{\beta}H +u_{\alpha}\tau_{\ab} + q_{\beta}$
\end{tabular}
\right]
= \left[ 
\begin{tabular}{c}
0 \\ $\mathcal{O}(\delta t)$ \\ $\mathcal{O}(\delta t)+3q$
\end{tabular}
\right],
\end{align}

where

\begin{align}
\tau_{\ab} = - \mu \left(\partial_{\alpha}u_{\beta}+\partial_{\beta}u_{\alpha} - \frac{2}{D}\partial_{\gamma}u_{\gamma}\delta_{\ab}\right)
- \eta \partial_{\gamma}u_{\gamma}\delta_{\ab},
\label{eq:shear-stress}
\end{align}

is the shear stress tensor and $q_{\beta} = -k\partial_{\beta}T$ is the Fourier heat flux.
The shear viscosity, bulk viscosity and the conductivity are
\begin{align}
\mu &= \left(\frac{1}{\omega}-\frac{1}{2}\right)p\delta t \label{eq:viscosity}, \\
\eta &=\left(\frac{1}{\omega}-\frac{1}{2}\right)\left(\frac{D+2}{D}-\gamma\right)p\delta t,  \\
k &=\left(\frac{1}{\omega}-\frac{1}{2}\right)\left(\frac{D-1}{2}\right)p\delta t,\label{eq:conductivity}
\end{align}
respectively. We note that  the bulk viscosity vanishes at the limit of 
a monatomic ideal-gas, as expected \cite{ansumali2005prl}.\\

The error term in the momentum and energy equations are found as
\begin{align}
&X_{\alpha M} = -\delta t \rho U_{\ab}\partial_{\beta}\theta,\\
&X_E = -\frac{\delta t}{2} \left( -\rho\frac{\theta}{C_v}(\partial_{\alpha}u_{\alpha})^2 - \partial_{\alpha}\theta\partial_{\alpha}(\rho\theta\Tref)
+ 4\rho u_{\alpha}\partial_{\beta}\theta U_{\beta\alpha} \right)\nonumber\\
&- 3\bar{q},
\label{eq:d2q9-error}
\end{align}
where
\begin{align}
\bm{U} = \bm{\nabla}\bm{u}\odot \bm{I} = \left[
\begin{tabular}{cc}
$\partial_x u$ & 0\\
0 & $\partial_y v$
\end{tabular}
\right],
\end{align} 
 is the Hadamard product of the velocity gradient tensor and the identity matrix and
\begin{align}
\bar{q} = \mu \left[ (3-\gamma)\left((\partial_x u)^2 + (\partial_y v)^2\right) + 2 (1-\gamma)\partial_x u \partial_y v\right].
\end{align} 
Finally, the Prandtl number is found as
\begin{align}
{\rm Pr} =\frac{\mu C_p}{k} =  \frac{D+2}{D-1},
\label{eq:prandtl}
\end{align}
which amounts to $4$ in two dimensions as reported in \cite{pond}.

\subsubsection{$D2Q25$}

The $D2Q25$ lattice is a tensor product of two $D1Q5$ lattices with the independent moment system of
\begin{align}
M_{mn} = \sum_i f_i v_{ix}^m v_{iy}^n,
\begin{tabular}{r}
$m=0,...,4$\\
$n=0,...,4$
\end{tabular}
\end{align}
where the property \eqref{eq:inverse-moment-prop} holds for the inversion mapping matrix $\mathcal{M}^{-1}$.
While the first-order equations coincide with those obtained in \eqref{eq:d2q9-first-order},
the second-order equations are derived as

\begin{align}
\dtt\left[
\begin{tabular}{c}
$\rho$ \\ $\rho u_{\alpha}$ \\ $P_{\alpha\alpha}^{\eq}$
\end{tabular}
\right]
+
\dbeta\left[
\begin{tabular}{c}
$0$ \\ $(1-\omega/2)P^{(1)}_{\ab}$ \\ $(1-\omega/2)Q^{(1)}_{\alpha\ab}$
\end{tabular}
\right]
= \left[ 
\begin{tabular}{c}
0 \\ 0\\ 0
\end{tabular}
\right].
\label{eq:d2q25-second-order}
\end{align}
The non-equilibrium pressure tensor is recovered as in \eqref{eq:noneq-pressure} however, the non-equilibrium heat flux is derived as
\begin{align}
Q_{\alpha\alpha\beta}^{(1)} = -\frac{\delta t}{\omega} \left[\dt Q_{\alpha\alpha\beta}^{\eq}
+ \dalpha R_{\ab}^{\eq}\right],
\end{align}

Finally, the hydrodynamic equations for the $D2Q25$ lattice are recovered as

\begin{align}
\partial_t\left[
\begin{tabular}{c}
$\rho$ \\ $\rho u_{\alpha}$ \\ $\rho E$
\end{tabular}
\right]
+
\partial_{\beta}\left[
\begin{tabular}{c}
$\rho u_{\beta}$ \\ $\rho u_{\alpha}u_{\beta} + p\delta_{\ab}+\tau_{\ab}$ \\ $\rho u_{\beta}H +u_{\alpha}\tau_{\ab} + q_{\beta}$
\end{tabular}
\right]
= \left[ 
\begin{tabular}{c}
0 \\ 0 \\ 0
\end{tabular}
\right],
\end{align}

where the shear stress tensor $\tau_{\ab}$ is defined in Eq. \eqref{eq:shear-stress} with the dynamic viscosity \eqref{eq:viscosity}.
The conductivity however, is recovered as

\begin{align}
k &=\left(\frac{1}{\omega}-\frac{1}{2}\right)p\delta t\left(\frac{D+2}{2}\right),
\label{eq:conductivity-25}
\end{align}

which implies ${\rm Pr} = 1$. 

Similar to the $D1Q5$ lattice, the hydrodynamic equations are recovered free of error terms.

\subsection{Variable specific heat}
\label{sec:variable-gamma}
To achieve an arbitrary specific heat $\gamma$, it is conventional to adopt a second population. However,
one can assign the second population to either carry the total energy or the extra internal energy.
We introduce the following equilibrium for the second population
\begin{align}
g_i^{\eq} = 2f_i^{\eq} \left[\left(C_v-\frac{D}{2}\right)T + (1-\phi)\frac{v_i^2}{2}\right],
\end{align}
where $\phi=1$ implies that the excess internal energy as the difference from a $D$ dimensional gas
is conserved by the $g$ population, while the kinetic energy is maintained by the $f$ population \cite{frapolli2016entropic}.
On the other hand, $\phi=0$ corresponds to the conservation of the total energy by the $g$ population.
Consequently, the total energy is computed as
\begin{align}
2\rho E = 2\rho e + \rho u^2 = \sum_i g_i + \phi \sum_i f_i v_i^2.
\end{align}
Since the hydrodynamic equations for the $D2Q25$ lattice are free of error terms,
it seems natural to choose $\phi=1$ so the equilibrium function of the second population 
is only a function of temperature and free of velocity terms. However, the choice of $\phi=1$ for 
the $D2Q9$ lattice will retain the error terms in the momentum and total energy equations, with the 
only difference that the specific heat will possess an arbitrary value instead of that of a monatomic 
ideal gas. In this case, the Prandtl number becomes

\begin{align}
{\rm Pr} = \frac{2\gamma}{3-\gamma},
\label{eq:pond/prandtl}
\end{align}
which limits the value of the adiabatic exponent to $\gamma=3$ as higher values 
will amount to unphysical answers. 
On the other hand, choosing $\phi=0$ will remove the errors from the total energy equation.
Since there will be no closure relation for the diagonal elements of the non-equilibrium third order moment,
the Prandtl number will take its natural value ${\rm Pr}=1$, independent of the choice of $\gamma$.
Nevertheless, a variable Prandtl number can always be achieved by using two relaxation parameters \cite{frapolli2016entropic}.

\subsection{Interpolation}
\label{sec:interpolation}
So far, the analysis have been carried out assuming a continuous space, whereas one must account for the
interpolation of transformed populations during the advection process. As mentioned before, the departure
point accessed during the semi-Lagrangian advection does not essentially coincide with a grid point and 
a set of collocation points are required to interpolate for the missing information.

\begin{figure}[!t]
\centering
\includegraphics[clip, trim = 1cm 2cm 0 1.5cm, width=\linewidth]{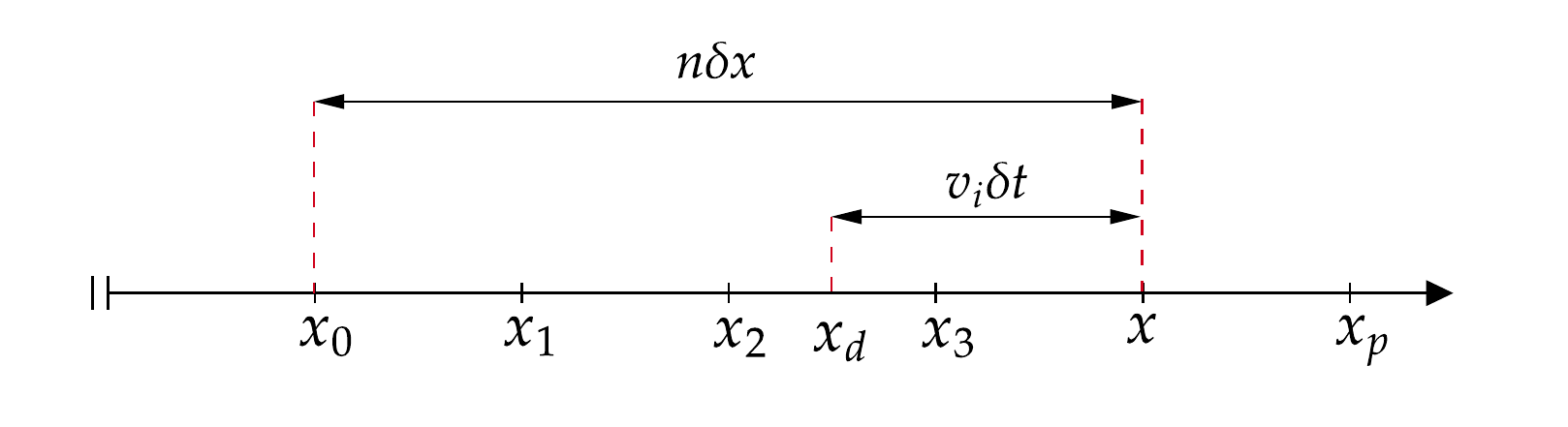}
\caption{Schematic of the interpolation process and the collocation points during the semi-Lagrangian advection.}
\label{fig:interpolation}
\end{figure}

In order to proceed with the analysis, we consider the discretized form \eqref{eq:SL-advection},
where $N$ number of points are used for the interpolation. Without loss of generality, we assume
$v_i > 0$. The departure point will be located on an off-grid point $x_d = x-v_i \delta t$, where $x$
is a grid point. Depending on the order of the interpolation, a set of grid points $x_p$ around the 
departure point will be used for the interpolation process. We assume that the first point of this 
stencil is located in the distance $n\delta x$ from the monitoring point $x$ such that
$x_0 = x-n\delta x$, where $n$ is an integer (see Fig. \ref{fig:interpolation}). 
The advecton equation \eqref{eq:SL-advection} is recast in the following form
\begin{align}
f_i^{\lambda}(x,t)=\sum_{p=0}^{N-1} l_p\mathcal{M}_{i,\lambda}^{-1} M^* (x_p,t-\delta t),
\label{eq:advection-interpolation}
\end{align}
where $l_p$ are the interpolation weights. \rev{At this point, no explicit type of the interpolation
function is assumed and the weights $l_p$ or their properties remain to be derived.}
Considering that $x_p = x-(n-p)\delta x$, one can expand Eq. \eqref{eq:advection-interpolation} using the
Taylor series up to third-order terms
\begin{align}
f_i(x,t) &= \sum_p l_p \mathcal{M}_i^{-1} \left(M^*(x,t)-\delta t \bar{D}_p M^*(x,t) + \frac{\delta t^2}{2}\bar{D}_p^2 M^*(x,t)\right) \nonumber\\
	 &= f_i^*(x,t)\sum_p l_p  -\delta t\mathcal{M}_i^{-1} \sum_p l_p \bar{D}_p M^*(x,t) \nonumber\\
         &+ \frac{\delta t^2}{2}\mathcal{M}_i^{-1}\sum_p l_p \bar{D}_p^2 M^*(x,t),
\label{eq:advection-interpolation-expanded}
\end{align}
where $\bar{D}_p = \partial_t + (n-p)(\delta x/\delta t)\partial_x;\ p=0:N-1$.\\

In the following, we shall compute the individual terms in Eq. \eqref{eq:advection-interpolation-expanded}.
In order to be consistent, any interpolation scheme requires the weights to sum to unity, i.e.
\begin{align}
\sum_p l_p = 1.
\end{align}
The other terms are computed as follows
\begin{align}
\sum_p l_p \bar{D}_p M^* = \bar{D}_0 M^*-\sum_p l_p p\frac{\delta x}{\delta t}\partial_x M^*,
\label{eq:term1}
\end{align}
\begin{align}
\sum_p l_p \bar{D}^2_p M^* &= \bar{D}^2_0 M^*+\left(\sum_p l_p p^2-2n\sum_p l_p p\right) \frac{\delta x^2}{\delta t^2}\partial_{xx} M^*\nonumber\\
& -2\sum_p l_p p \frac{\delta x}{\delta t} \partial_{xt}M^*,
\label{eq:term2}
\end{align}
In a moment-conserving interpolation function \cite{koumoutsakos1997inviscid,rees20143d}, we have the property
\begin{align}
\sum_p l_p (p\delta x)^r = (x_d-x_0)^r,
\label{eq:moment-conserving}
\end{align}
where the number of conserved moments $r$ depends on the order of the interpolation function.
Hence, we require the interpolation scheme to obey the property \eqref{eq:moment-conserving} with
$r=2$ at least. This implies that a stencil with a minimum of three points must be used for the interpolation.

Substituting Eq. \eqref{eq:moment-conserving} in Eqs. \eqref{eq:term1} and \eqref{eq:term2} leads to
\begin{align}
\sum_p l_p \bar{D}_p M^* = \partial_t M^* + v_i\partial_x M^*=D_i M^*,
\label{eq:term1-new}
\end{align}

\begin{align}
\sum_p l_p \bar{D}^2_p M^* &= \partial_{tt}M^* + 2v_i \partial_{xt}M^* + v_i^2\partial_{xx}M^* \nonumber\\
&= D^2_i M^*-D_i v_i \partial_x M^*.
\label{eq:term2-new}
\end{align}

It can be simplify verified that once the averaged terms \eqref{eq:term1-new} and \eqref{eq:term2-new} are
plugged in Eq. \eqref{eq:advection-interpolation-expanded}, the kinetic equation of the continuous case 
\eqref{eq:evolution-f-1} is recovered. Therefore, all the analysis presented so far are also valid when the interpolation 
procedure is included provided that the interpolation function encompasses three support points at least and
abides the  moment-conserving property.

\subsection{Prandtl number}

In section \ref{sec:variable-gamma}, it was shown that the choice of equilibrium for the second population for a standard lattice such as $D2Q9$
will affect the recovered energy equation. Beside the unwanted error terms, the Prandtl number obtained by the Chapman-Enskog analysis
will be a rational function of the specific heat, i.e. Eq. \eqref{eq:pond/prandtl}, if only the extra internal energy
is assigned to the second population $(\phi=1)$. On the other hand, choosing $\phi=0$ will remove the error terms 
and recover ${\rm Pr} = 1$. Moreover, we illustrated that in order to have a consistent scheme, the interpolation function
must feature a moment conserving property with at least three support points. 

\begin{figure}[!t]
\centering
\includegraphics[clip, trim = 0 0 0 1.0cm, width=\linewidth]{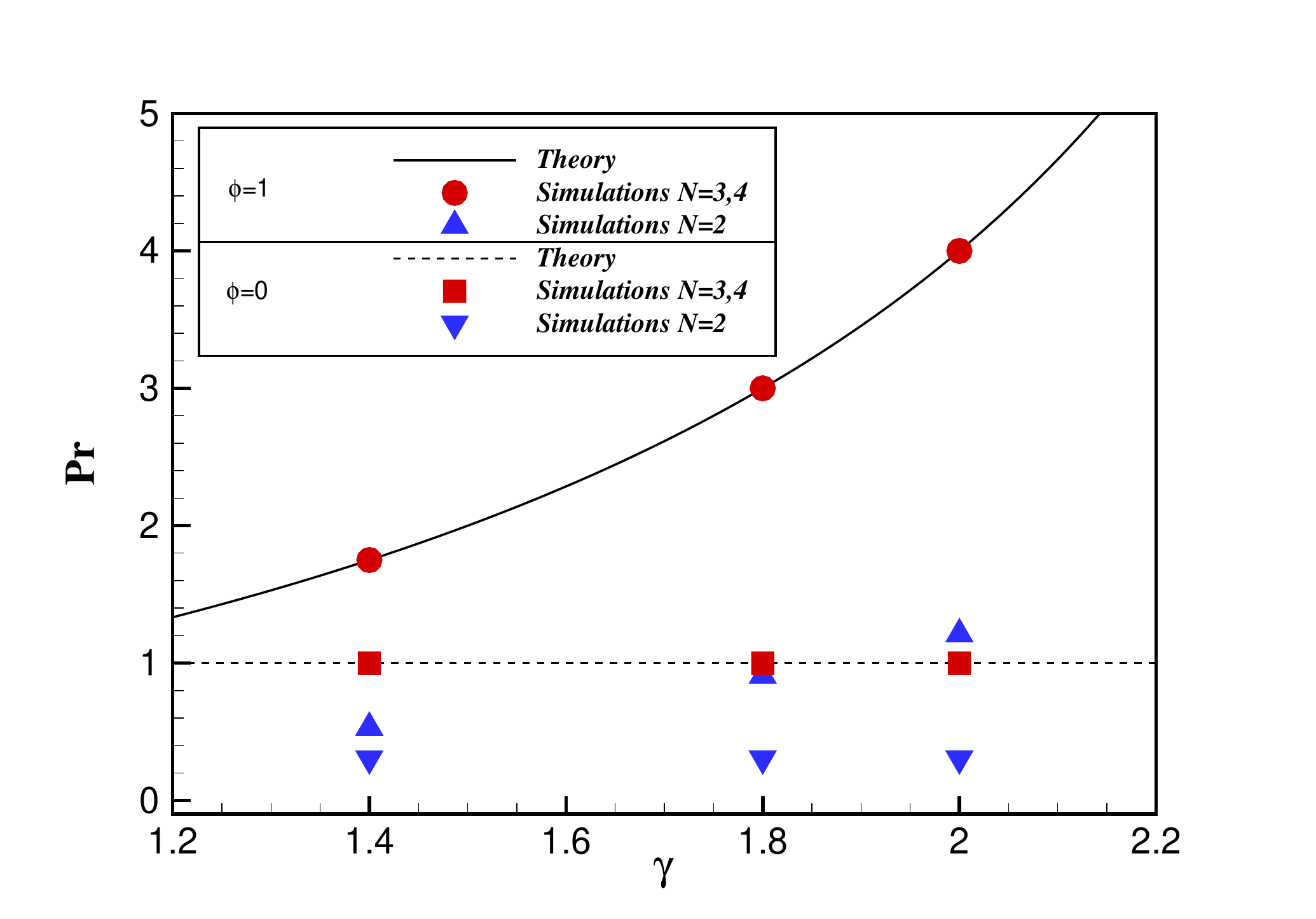}
\caption{The measured Prandtl number against the adiabatic exponent using the $D2Q9$ lattice and different values of $\phi$.}
\label{fig:pond/prandtl}
\end{figure}

To verify our analysis, we conduct the standard test case
to measure the value of the Prandtl number \cite{pond}. We choose the $D2Q9$ lattice with the
first-order $(N=2, r=1)$, second-order $(N=3, r=2)$ and third-order $(N=4, r=3)$ Lagrange interpolation schemes.
Figure \ref{fig:pond/prandtl} shows that our analysis are consistent with the simulations. It is also clearly visible that the 
interpolation scheme with $r<2$ deviates from the underlying theoretical values.

\subsection{Convergence study}

The standard LBM is a second-order accurate scheme in space and time featuring $\delta x=\delta t=1$.
On the other hand, it is well-known that the compressibility errors in the standard LBM scale with ${\rm Ma}^2$
\rev{and} the NSE is recovered with error terms proportional to ${\rm Kn}^2$, where $\rm Kn$ is the Knudsen number \cite{kruger2017lattice}.
However, it has been shown that the semi-Lagrangian LBM (SLLBM) \cite{kramer2017semi,wilde2020semi}
can achieve higher orders by decoupling the time step from the grid spacing 
provided that the time step (or CFL) and the Mach number
is kept relatively low. Then, high-order interpolation functions can lead to 
high spatial order of accuracy. In this case, as shown and discussed in \cite{wilde2020semi,kramer2020lattice},
the discretization errors are in the order of $\mathcal{O}({\rm min}(\delta x^N/\delta t,\delta x^{N-1}))$.
On the other hand, we have shown that using 9 discrete velocities in the PonD framework will introduce error terms
in the order of $\mathcal{O}(\delta t)$ in the momentum and energy equations. Finally, one can summarize that the 
present model include error terms as deviations from the full compressible NSE, which are in the order of 
\begin{align}
D1Q3, D2Q9:& \nonumber\\
&\mathcal{O}\left({\rm min}\left(\frac{\delta x^N}{\delta t}, \delta x^{N-1}\right), \delta t, {\rm Kn}^2\right),\nonumber\\
D1Q5, D2Q25:&\nonumber\\
&\mathcal{O}\left({\rm min}\left(\frac{\delta x^N}{\delta t}, \delta x^{N-1}\right), \delta t^2, {\rm Kn}^2\right),
\label{eq:errors}
\end{align}
where the compressibility error $\mathcal{O}(\rm Ma^2)$ is eliminated thanks to the exact equilibrium function.
In the following, we will assess the validity of these results by conducting numerical simulations.
To verify the spatial discretization errors, a density profile is advected with the following initial conditions
\begin{align}
&\rho_0(x) = 1 + \exp(-300(x-0.5)^2),\ 0\leqslant x \leqslant 1, \nonumber \\
&p_0(x) = 1, \nonumber\\
&u_0(x) = 1.
\end{align}
While the number of grid points \rev{$Nx$} are varied in this simulation, the time step is fixed at a small value 
$\delta t=10^{-4}$ to eliminate the chance of dominance of temporal errors.
A third-order Lagrange interpolation function with four support points is adopted 
in this simulation and the value of the specific heat is chosen as $\gamma=1.4$.
We let the simulations run until $t=1$, which corresponds to one period in time.
\rev{To reflect the maximum error throughout the domain, the $L_{\infty}$-error defined as 
\begin{align}
L_{\infty}=\max\left(\left|\frac{\rho(x)-\rho_{0}(x)}{\rho_{0}(x)}\right|\right)
\end{align}
is measured to investigate the error convergence.}

\begin{figure}
\centering
\includegraphics[clip, trim = 0 0 0 1cm, width=\linewidth]{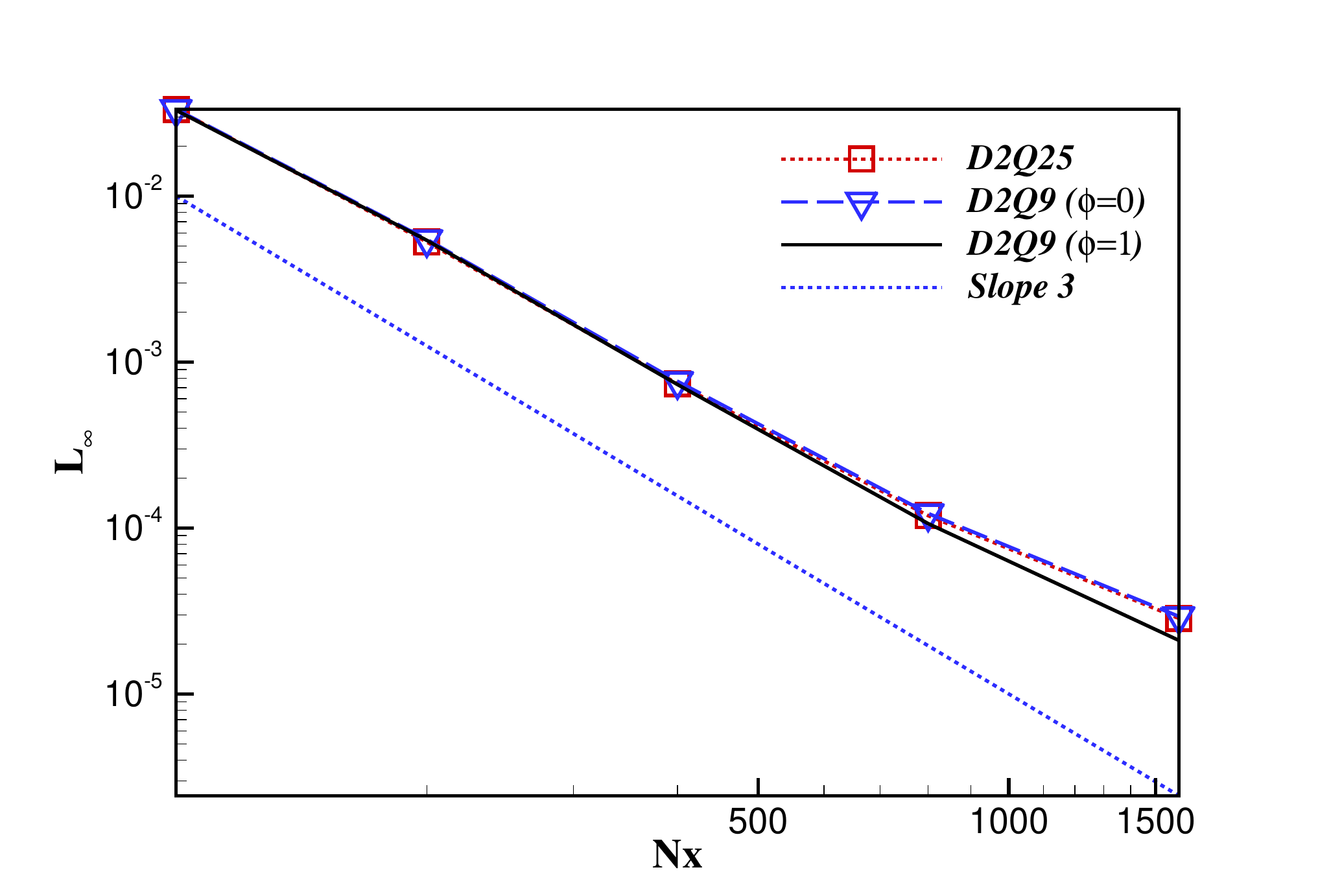}
	\caption{\rev{Convergence of the $L_{\infty}$ error in the linear advection test: all schemes recover the underlying order of accuracy.}}
\label{fig:pond/ovs}
\end{figure}

Figure \ref{fig:pond/ovs} shows that the 
underlying order of accuracy of the interpolation function is recovered for
both $D2Q25$ and $D2Q9$ lattices and is independent of the choice of the equilibrium function for the 
second population, as expected.

\begin{figure}
\centering
\includegraphics[clip, trim = 0 0 0 1cm, width=\linewidth]{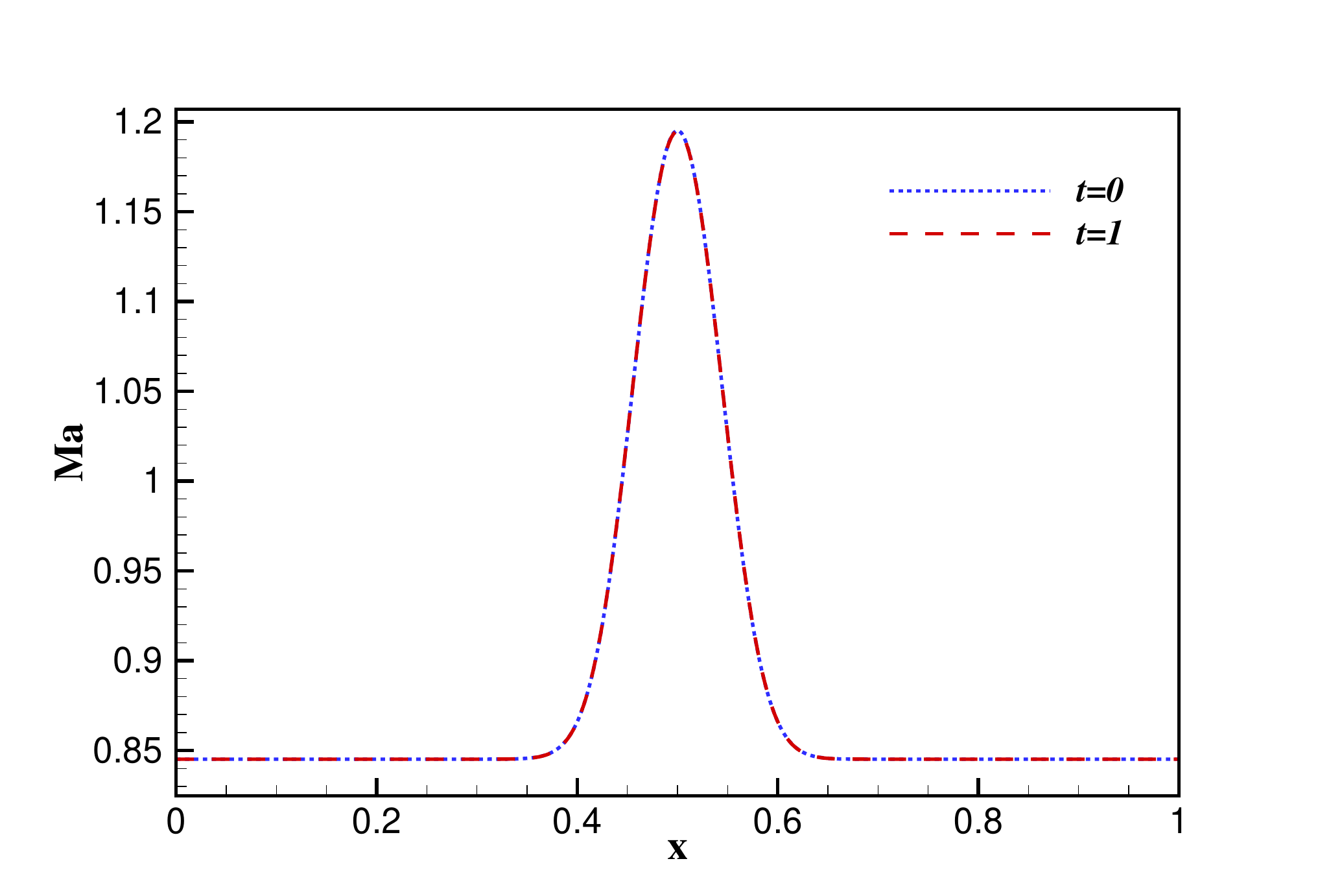}
	\caption{Mach profile of the linear advection test at $\delta t=10^{-4}$, using the $D2Q9$ lattice, $\rev{Nx}=400$ and $\phi=0$.}
\label{fig:pond/mach}
\end{figure}

Figure \ref{fig:pond/mach} shows the local Mach number for number of grid points $N=400$. 
It is noticed that the range of the Mach number in this simulation is significantly high, whereas
it was shown that the compressibility errors in SLLBM \cite{kramer2017semi} can already prevail at ${\rm Ma} = 0.1$.
This is due to the Galilean-invariant nature of the PonD model where it eliminates the compressibility errors by
designing particles at the co-moving reference frame and the exact collision seen from those particles

To study the behavior of the temporal errors, we simulate the advection of a vortex by a uniform flow
at ${\rm Ma}_a = U_a/\sqrt{\gamma T_0}=0.845$ \cite{Frapolli2016}. The velocity field in the cylindrical coordinates
and in the advected reference frame is $u_{\theta}(r)=u_{\rm max}r\exp[(1-r^2)/2]$, where $r=r^{\prime}/R$ is the reduced
radius and $R$ is the radius of the vortex. The vortex Mach number is defined based on the maximum tangential velocity
in the co-moving reference frame, ${\rm Ma}_v=u_{\rm max}/\sqrt{\gamma T_0}$ and is fixed to ${\rm Ma}=0.4$. 
The Reynolds number is fixed to ${\rm Re}=2U_a R/\nu=6\times 10^{5}$ and a $400\times400$ grid is used.
The vortex is allowed to complete one cycle of rotation during one period of advection and then the
x-velocity component is measured along the centerline, where its deviation from the exact solution
is indicative of errors. This simulation is repeated with different timesteps with a fixed advection velocity.
  
\begin{figure}
\centering
\includegraphics[clip, trim = 0 0 0 1cm, width=\linewidth]{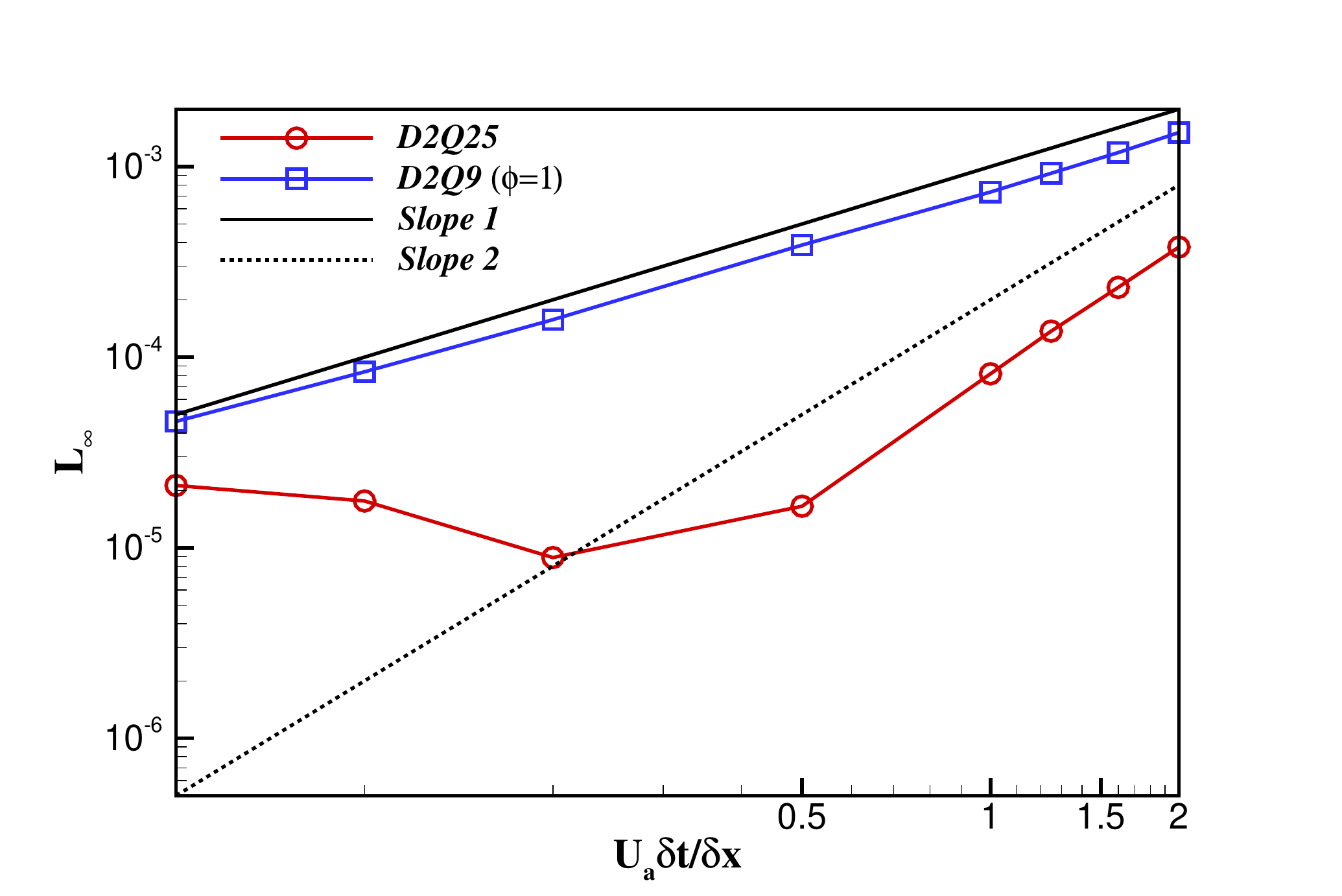}
	\caption{Time refinement study \rev{in the simulation of advected vortex: 
	convergence of the $L_{\infty}$ error by decreasing the timestep. $400\times400$ grid points are used.} }
\label{fig:pond/ovs-time}
\end{figure}

Figure \ref{fig:pond/ovs-time} shows the \rev{$L_{\infty}$} errors for both the $D2Q25$ and $D2Q9$ lattices.
We see the results are recovered consistently with Eq. \eqref{eq:errors}, where the $D2Q25$ lattice
shows second-order convergence, while the $D2Q9$ lattice is first-order in time when the excess internal energy
is assigned to the $g$ population, i.e. $\phi=1$. Another interesting point\rev{, which rises} in this simulation is the 
non-monotonic behavior of the temporal errors in the $D2Q25$ lattice. This is due to the \rev{competing effect between the 
$\mathcal{O}(\delta x^N/\delta t)$ and $\mathcal{O}(\delta t^2)$ error terms, when the resolution $\delta x$ and the order of the interpolation $N$ are fixed.
Depending on their orders of magnitude, the latter might take over at small time steps
and increase the errors as time is refined. At first, a second-order convergence is observed until $U_a \delta t/\delta x = 0.5$.
After this point, further refinement results in 
increasing the errors implying that the $\mathcal{O}(\delta x^N/\delta t)$ term has become dominant.}
This reverse effect, which was also reported in \cite{wilde2020semi} is not 
observed in the $D2Q9$ lattice in this setup since the $\mathcal{O}(\delta t)$ terms retain greater magnitude
throughout the refinement procedure.\\

\begin{figure}
\centering
\includegraphics[clip, trim = 0 0 0 1cm, width=\linewidth]{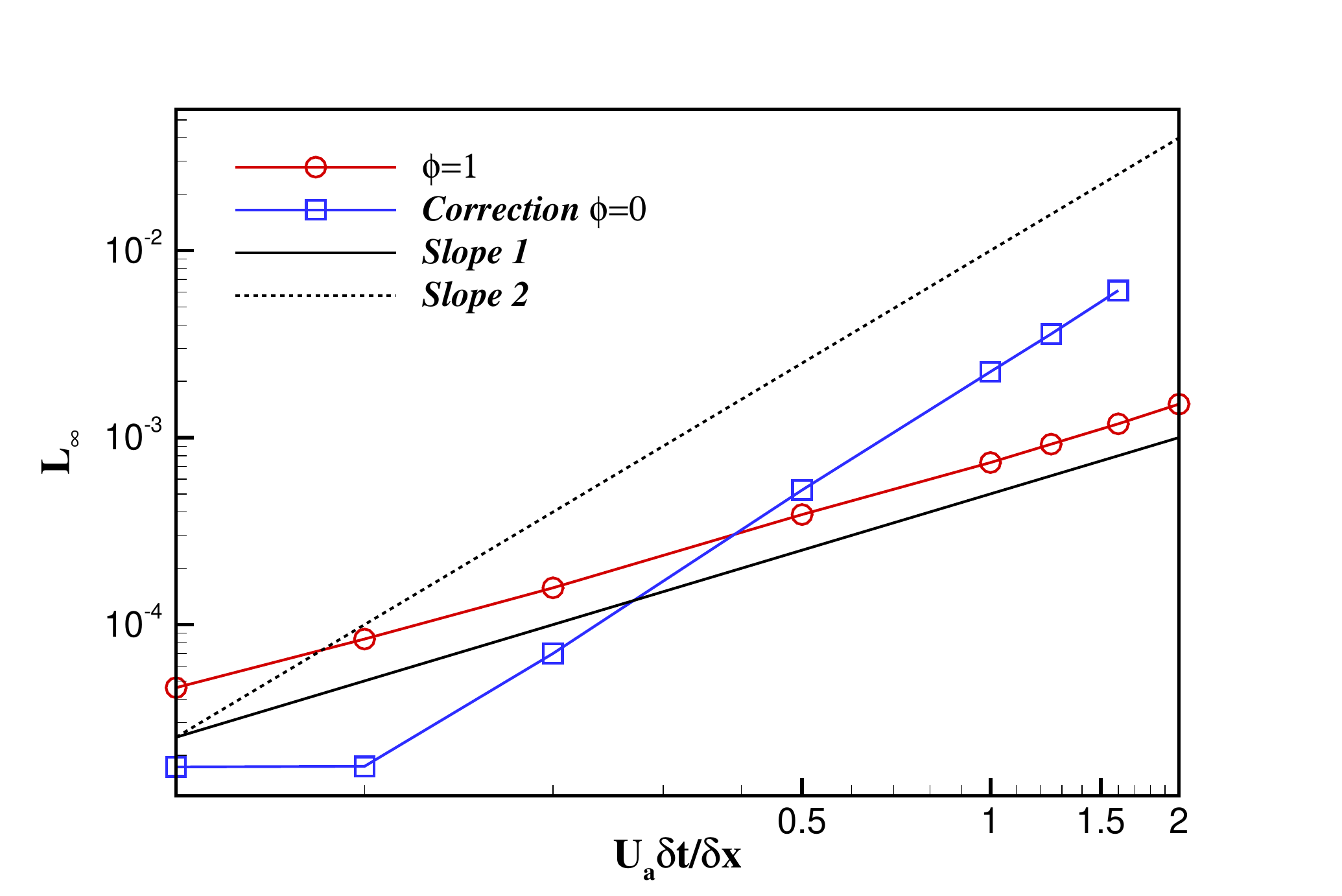}
	\caption{\rev{Time refinement study in the simulation of advected vortex: convergence of the $L_{\infty}$ error 
	by decreasing the timestep. The $D2Q9$ lattice is used with $\phi=1$ and $\phi=0$ augmented
	by the correction term.}}
\label{fig:ovs-time-corr}
\end{figure}

As said earlier, the choice of $\phi$ can affect the recovered energy equation when
the $D2Q9$ lattice is used, such that $\phi=0$ can remove
the errors from the energy equation. However, the momentum equation will still have the $\mathcal{O}(\delta t)$
errors and the scheme will be effectively first-order in time. To verify this, we augment the post-collision $f$ populations
by a forcing term as
\begin{align}
f_i^*(\x,t) = f_i(\x,t) + \omega(\rho w_i - f_i)_{(\x,t)} + \hat{f}_i\delta t,
\end{align}
where $\hat{f}_i = \mathcal{M}^{-1}_i\bm{X}$ and
\begin{align}
\bm{X}=[0,X_{xM},X_{yM},0,0,0,0,0,0]^{\dagger},
\label{eq:}
\end{align}
includes the error terms in the momentum equation recovered in \eqref{eq:d2q9-error}.
Choosing $\phi=0$ and forcing out the error terms $X_{\alpha M}$, we repeat the same simulation using the $D2Q9$ lattice.
As demonstrated in Fig. \ref{fig:ovs-time-corr}, the scheme becomes second-order accurate in time once the error terms are
corrected. This indicates that the analysis are consistent with simulations.

\section{Benchmark}
\begin{figure}[!t]
\centering
\includegraphics[clip, trim = 0 0 0 1cm, width=\linewidth]{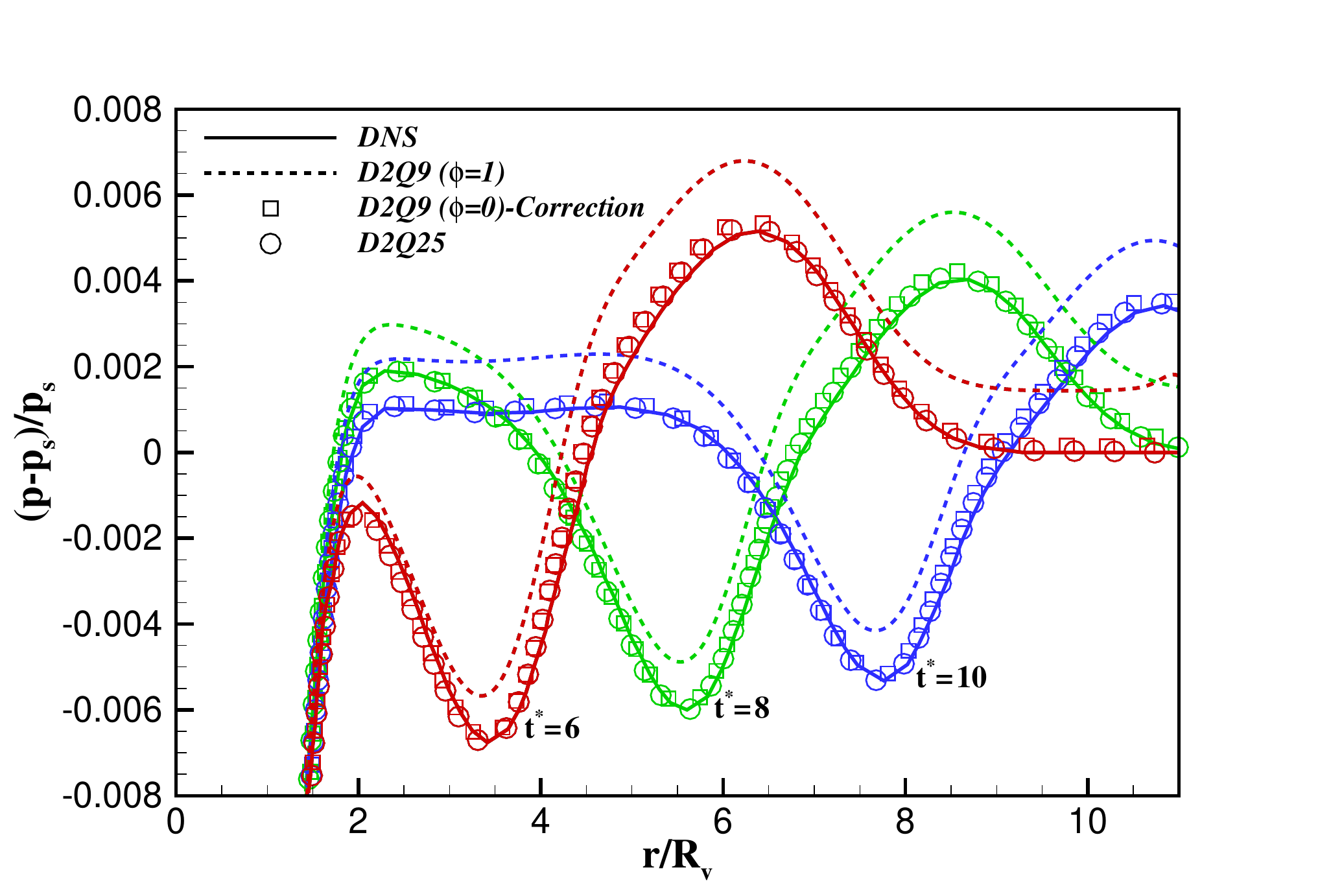}
\caption{Radial sound pressure measured at $\theta=-45^{\circ}$ with respect to the $x$ axis at nondimensional times $t^*=\{6, 8, 10\}$.
The governing parameters are ${\rm Ma}_a = 1.2, {\rm Ma}_v = 0.25, {\rm Re}=800$ and ${\rm Pr} = 0.75$.
DNS from \cite{inoue_hattori_1999}.} 
\label{fig:shock-vortex}
\end{figure}

In this section, the interaction of a vortex with a standing shock front is considered.
The Mach number of the vortex is ${\rm Ma}_v=0.25$ and its radius is denoted by $R_v$. When passing through
the shock front with the intensity ${\rm Ma}_a=1.2$, sound waves are generated by the vortex.
To assess the numerical accuracy of a model, one can measure the sound pressure
and compare to the DNS solution \cite{inoue_hattori_1999}. In this simulation, the Reynolds number is defined as
${\rm Re} = a_{\infty}R_v/\nu$, where $a_{\infty}$ is the speed of sound  upstream of the shock and
the dimensionless time $t^*=ta_{\infty}/R_v$ is used.
Figure \ref{fig:shock-vortex} shows the radial sound pressure measured from the center
of the vortex along the $\theta=-45^{\circ}$ line with respect to the $x$ axis. The results are
captured at three different times for
${\rm Re}=800$ and ${\rm Pr}=0.75$. 
Compared to the DNS solution, one observes that the $D2Q9$ lattice overestimates the pressure, while 
removing the error terms ($\phi=0$ and adding the correction term) leads to a significant improvement such that
the latter together with the $D2Q25$ lattice are in good agreement with the DNS solution.
\section{Conclusion}

In this paper, a consistent analysis of the particles on demand kinetic \rev{model} was presented.
Due to the off-lattice property of the model, the semi-Lagrangian advection is used, which requires
interpolation schemes. In our Chapman-Enskog analysis, we have taken into account the effect of the 
interpolation and transformation of populations during the advection process. By doing so, we have derived the
hydrodynamic limit of the model for commonly used one and two-dimensional lattices. 
It has been demonstrated that the $D2Q9$ lattice in the PonD framework have error terms 
in the order of timestep in the momentum and energy equations, while the $D2Q25$ lattice can
recover the full compressible NSE. However, the error terms corresponding to the energy equation 
recovered from the $D2Q9$ lattice could be eliminated by adopting a second population to carry the total 
energy instead of the excess internal energy. 

Furthermore, we discussed that similar to other semi-Lagrangian LB schemes, the
spatial order of accuracy can be increased by employing high-order interpolation functions
at low {\rm CFL} numbers. However, the compressibility errors are no longer present thanks to the
Galilean-invariant nature of the model. 

\rev{As for the validation,} the presented analysis were verified on various numerical benchmarks.
It was shown that the results were improved upon including corrections to remove the error terms.

\rev{Finally, we comment that the current analysis can be applied to three-dimesnional lattices.
In the case of tensor-product lattices such as $D3Q27$ and $D3Q125$, the recovered equations for 
fluid properties
such as viscosity, conductivity and Prandtl number 
 will be the same as their two-dimensional counterpart (Eqs. \eqref{eq:viscosity}-\eqref{eq:conductivity}, \eqref{eq:prandtl},
\eqref{eq:conductivity-25}).
However, for any general lattice, a separate investigation must be carried out due to their unique form 
of mapping matrix.}

\section{Acknowledgement}
The author thanks I. Karlin and B. Dorschner for the discussions.
\rev{ 
\appendix
\section{}
In this section, we aim to analyze the predictor-corrector step during the advection process.
To this end, we take the $D1Q3$ lattice with the inverted mapping matrix $\mathcal{M}^{-1}$ as shown
by Eq. \eqref{eq:mapping-inverse}. For simplicity, the interpolation kernel is once again neglected.
The goal is to find the co-moving gauge at point $x$ and time $t_0+\delta t$ assuming that the flow 
variables are  known at time $t_0$. 
As explained in section \ref{sec:pond/kinetic-eq}, the particles at point $x$ are first set relative to
the initial gauge $\lambda_0=\lambda(x,t_0)$. Hence the initial value of the discrete velocities are
\begin{align}
v_i^0 = v_i(x,t_0)=\sqrt{\theta_0}c_i + u_0,
\end{align} 
where $\theta_0 = \theta(x,t_0)$ and $u_0 = u(x,t_0)$.
At the next step, the semi-Lagrangian advection is followed using Eq. \eqref{eq:streaming-recast}
\begin{align}
f_i^{\lambda_0} = \mathcal{M}_{i,\lambda_0}^{-1} M^*(x-v_i^0\delta t,t_0).
\end{align}
Consequently, using Eqs. \eqref{eq:macro-moments-rho}-\eqref{eq:macro-moments-temp},
the updated density, momentum and total energy values are obtained and denoted as 
$\rho_1, (\rho u)_1, (2\rho E)_1$, respectively.\\
In general, one can express the populations at the $n$th iteration by
\begin{align}
f_i^{\lambda_n} = \mathcal{M}_{i,\lambda_n}^{-1} M^*(x-v_i^n\delta t,t_0),
\end{align}
where $\lambda_n=\{T_n,u_n\}$ denotes the reference frame corresponding to the computed temperature 
$T_n=\theta_n\Tref$ and velocity $u_n$. The flow variables are updated as
\begin{align}
\rho_{n+1} &= \sum_i      f_i^{\lambda_n},\\
(\rho u)_{n+1} &= \sum_i  f_i^{\lambda_n} v_i^n,\\
(2\rho E)_{n+1} &= \sum_i f_i^{\lambda_n} {v_i^n}^2.
\end{align}
It is straightforward to show 
\begin{align}
f_0^{\lambda_n} &= \left(1-\frac{u_n^2}{\theta_n}\right)\rho(x_0^n) + \frac{2u_n}{\theta_n}(\rho u)(x_0^n)-\frac{1}{\theta_n}(2\rho E)(x_0^n),\nonumber\\
f_1^{\lambda_n} &= \frac{u_n^2-\sqrt{\theta_n}u_n}{2\theta_n}\rho(x_1^n) + \frac{\sqrt{\theta_n}-2u_n}{2\theta_n}(\rho u)(x_1^n)+\frac{1}{2\theta_n}(2\rho E)(x_1^n),\nonumber\\
f_2^{\lambda_n} &= \frac{u_n^2+\sqrt{\theta_n}u_n}{2\theta_n}\rho(x_2^n) - \frac{\sqrt{\theta_n}+2u_n}{2\theta_n}(\rho u)(x_2^n)+\frac{1}{2\theta_n}(2\rho E)(x_2^n),
\label{appendix:populations}
\end{align}
where $x_j^n; j=0,1,2$ are the departure points as shown in Fig. \ref{fig:streaming}, at the $n$th iteration, such that
\begin{align}
x_0^n &= x - u_n\delta t, \nonumber\\
x_1^n &= x - \sqrt{\theta_n}\delta t - u_n\delta t, \nonumber\\
x_2^n &= x + \sqrt{\theta_n}\delta t - u_n\delta t.
\end{align}
It can be observed that $x_0^n$ is always the middle point between $x_1^n$ and $x_2^n$.
Note that the moments $[\rho,\rho u,2\rho E]$ in Eq. \eqref{appendix:populations} -- evaluated at
departure points $x_j^n$ -- are known from the previous time step $t=t_0$. 
\blue{It must be commented that the transformation process is not positivity preserving, i.e. 
the transformed populations can assume negative values depending on the target reference frame (see Eq. \eqref{appendix:populations}}.\\
Before proceeding further, we use Taylor series
to expand these terms around $(x,t_0)$ up to third order accuracy.
Finally, taking the first two moments of \ref{appendix:populations}, we have
\begin{align}
	\rho_{n+1}     &=  \rho   (x,t_0) - \delta t\partial_x(\rho u) + \frac{\delta t^2}{2}\partial_{xx}(2\rho E)+\mathcal{O}(\delta t^3),\\
	(\rho u)_{n+1} &= (\rho u)(x,t_0) + \frac{\delta t^2}{2}u_n^3\partial_{xx}\rho-\frac{\delta t^2}{2}u_n\theta_n\partial_{xx}\rho \nonumber\\
	&- \frac{3}{2}\delta t^2u_n^2\partial_{xx}(\rho u) + \frac{\delta t^2}{2}\theta_n\partial_{xx}(\rho u)
	 - \delta t\partial_x(2\rho E) \nonumber\\
	 &+ \frac{3}{2}\delta t^2 u_n\partial_{xx}(2\rho E) \label{appendix:momentum}
\end{align}
It is observed that the evaluated density during the iterations is independent of its previous values (there is no dependence on $\rho_n$).
Hence, one can write $\rho=\rho_n=\rho_{n+1}$.
For further simplification, assume the isothermal condition, i.e., $\theta_n=1$. 
According to Eq. \eqref{appendix:momentum}, the difference of computed momentums between two subsequent iterations is  
\begin{align}
	\rho (u_{n+1}- u_{n})&= \frac{\delta t^2}{2}(u_n^3-u_{n-1}^3)\partial_{xx}\rho-\frac{\delta t^2}{2}(u_n-u_{n-1})\partial_{xx}\rho \nonumber\\
	&- \frac{3}{2}\delta t^2(u_n^2-u_{n-1}^2)\partial_{xx}(\rho u)\nonumber\\ 
	 &+ \frac{3}{2}\delta t^2 (u_n-u_{n-1})\partial_{xx}(2\rho E).
\label{appendix:momentum-2}
\end{align}
It is straightforward to show that
\begin{align}
	\left|\frac{u_{n+1}- u_{n}}{u_n-u_{n-1}}\right|= \mathcal{O}(\delta t^2).
\end{align}
Hence, the predictor-corrector algorithm is always convergent at relatively low time steps.}\\

\rev{Another approach is to evaluate the derivative of the iteration function.
According to the fixed-point iteration method, the iterative process $x_{n+1}=\Psi(x_n)$ will be convergent if
$|\Psi^{\prime}(x)|<1$ for a specified interval.\\
Rearranging Eq. \eqref{appendix:momentum-2} as $u_{n+1}=\Psi(u_n)$, it can be shown that
\begin{align}
\left| \frac{\partial \Psi}{\partial u}\right| = \mathcal{O}(\delta t^2).
\end{align}
}


\begin{thebibliography}{27}%
\makeatletter
\providecommand \@ifxundefined [1]{%
 \@ifx{#1\undefined}
}%
\providecommand \@ifnum [1]{%
 \ifnum #1\expandafter \@firstoftwo
 \else \expandafter \@secondoftwo
 \fi
}%
\providecommand \@ifx [1]{%
 \ifx #1\expandafter \@firstoftwo
 \else \expandafter \@secondoftwo
 \fi
}%
\providecommand \natexlab [1]{#1}%
\providecommand \enquote  [1]{``#1''}%
\providecommand \bibnamefont  [1]{#1}%
\providecommand \bibfnamefont [1]{#1}%
\providecommand \citenamefont [1]{#1}%
\providecommand \href@noop [0]{\@secondoftwo}%
\providecommand \href [0]{\begingroup \@sanitize@url \@href}%
\providecommand \@href[1]{\@@startlink{#1}\@@href}%
\providecommand \@@href[1]{\endgroup#1\@@endlink}%
\providecommand \@sanitize@url [0]{\catcode `\\12\catcode `\$12\catcode
  `\&12\catcode `\#12\catcode `\^12\catcode `\_12\catcode `\%12\relax}%
\providecommand \@@startlink[1]{}%
\providecommand \@@endlink[0]{}%
\providecommand \url  [0]{\begingroup\@sanitize@url \@url }%
\providecommand \@url [1]{\endgroup\@href {#1}{\urlprefix }}%
\providecommand \urlprefix  [0]{URL }%
\providecommand \Eprint [0]{\href }%
\providecommand \doibase [0]{http://dx.doi.org/}%
\providecommand \selectlanguage [0]{\@gobble}%
\providecommand \bibinfo  [0]{\@secondoftwo}%
\providecommand \bibfield  [0]{\@secondoftwo}%
\providecommand \translation [1]{[#1]}%
\providecommand \BibitemOpen [0]{}%
\providecommand \bibitemStop [0]{}%
\providecommand \bibitemNoStop [0]{.\EOS\space}%
\providecommand \EOS [0]{\spacefactor3000\relax}%
\providecommand \BibitemShut  [1]{\csname bibitem#1\endcsname}%
\let\auto@bib@innerbib\@empty
\bibitem [{\citenamefont {Succi}(2018)}]{succi2018lattice}%
  \BibitemOpen
  \bibfield  {author} {\bibinfo {author} {\bibfnamefont {S.}~\bibnamefont
  {Succi}},\ }\href@noop {} {\emph {\bibinfo {title} {The lattice Boltzmann
  equation: for complex states of flowing matter}}}\ (\bibinfo  {publisher}
  {Oxford University Press},\ \bibinfo {year} {2018})\BibitemShut {NoStop}%
\bibitem [{\citenamefont {Kr{\"u}ger}\ \emph {et~al.}(2017)\citenamefont
  {Kr{\"u}ger}, \citenamefont {Kusumaatmaja}, \citenamefont {Kuzmin},
  \citenamefont {Shardt}, \citenamefont {Silva},\ and\ \citenamefont
  {Viggen}}]{kruger2017lattice}%
  \BibitemOpen
  \bibfield  {author} {\bibinfo {author} {\bibfnamefont {T.}~\bibnamefont
  {Kr{\"u}ger}}, \bibinfo {author} {\bibfnamefont {H.}~\bibnamefont
  {Kusumaatmaja}}, \bibinfo {author} {\bibfnamefont {A.}~\bibnamefont
  {Kuzmin}}, \bibinfo {author} {\bibfnamefont {O.}~\bibnamefont {Shardt}},
  \bibinfo {author} {\bibfnamefont {G.}~\bibnamefont {Silva}}, \ and\ \bibinfo
  {author} {\bibfnamefont {E.~M.}\ \bibnamefont {Viggen}},\ }\href@noop {}
  {\bibfield  {journal} {\bibinfo  {journal} {Springer International
  Publishing}\ }\textbf {\bibinfo {volume} {10}},\ \bibinfo {pages} {4}
  (\bibinfo {year} {2017})}\BibitemShut {NoStop}%
\bibitem [{\citenamefont {Sbragaglia}\ \emph {et~al.}(2006)\citenamefont
  {Sbragaglia}, \citenamefont {Benzi}, \citenamefont {Biferale}, \citenamefont
  {Succi},\ and\ \citenamefont {Toschi}}]{Sbragaglia2006}%
  \BibitemOpen
  \bibfield  {author} {\bibinfo {author} {\bibfnamefont {M.}~\bibnamefont
  {Sbragaglia}}, \bibinfo {author} {\bibfnamefont {R.}~\bibnamefont {Benzi}},
  \bibinfo {author} {\bibfnamefont {L.}~\bibnamefont {Biferale}}, \bibinfo
  {author} {\bibfnamefont {S.}~\bibnamefont {Succi}}, \ and\ \bibinfo {author}
  {\bibfnamefont {F.}~\bibnamefont {Toschi}},\ }\href {\doibase
  10.1103/PhysRevLett.97.204503} {\bibfield  {journal} {\bibinfo  {journal}
  {Physical Review Letters}\ }\textbf {\bibinfo {volume} {97}},\ \bibinfo
  {pages} {204503} (\bibinfo {year} {2006})}\BibitemShut {NoStop}%
\bibitem [{\citenamefont {Biferale}\ \emph {et~al.}(2012)\citenamefont
  {Biferale}, \citenamefont {Perlekar}, \citenamefont {Sbragaglia},\ and\
  \citenamefont {Toschi}}]{Biferale2012}%
  \BibitemOpen
  \bibfield  {author} {\bibinfo {author} {\bibfnamefont {L.}~\bibnamefont
  {Biferale}}, \bibinfo {author} {\bibfnamefont {P.}~\bibnamefont {Perlekar}},
  \bibinfo {author} {\bibfnamefont {M.}~\bibnamefont {Sbragaglia}}, \ and\
  \bibinfo {author} {\bibfnamefont {F.}~\bibnamefont {Toschi}},\ }\href
  {\doibase 10.1103/PhysRevLett.108.104502} {\bibfield  {journal} {\bibinfo
  {journal} {Physical Review Letters}\ }\textbf {\bibinfo {volume} {108}},\
  \bibinfo {pages} {104502} (\bibinfo {year} {2012})},\ \Eprint
  {http://arxiv.org/abs/1111.0905} {arXiv:1111.0905} \BibitemShut {NoStop}%
\bibitem [{\citenamefont {Benzi}\ \emph {et~al.}(2009)\citenamefont {Benzi},
  \citenamefont {Chibbaro},\ and\ \citenamefont {Succi}}]{Benzi2009}%
  \BibitemOpen
  \bibfield  {author} {\bibinfo {author} {\bibfnamefont {R.}~\bibnamefont
  {Benzi}}, \bibinfo {author} {\bibfnamefont {S.}~\bibnamefont {Chibbaro}}, \
  and\ \bibinfo {author} {\bibfnamefont {S.}~\bibnamefont {Succi}},\ }\href
  {\doibase 10.1103/PhysRevLett.102.026002} {\bibfield  {journal} {\bibinfo
  {journal} {Physical Review Letters}\ }\textbf {\bibinfo {volume} {102}},\
  \bibinfo {pages} {026002} (\bibinfo {year} {2009})}\BibitemShut {NoStop}%
\bibitem [{\citenamefont {{Mazloomi M}}\ \emph {et~al.}(2015)\citenamefont
  {{Mazloomi M}}, \citenamefont {Chikatamarla},\ and\ \citenamefont
  {Karlin}}]{MazloomiM2015}%
  \BibitemOpen
  \bibfield  {author} {\bibinfo {author} {\bibfnamefont {A.}~\bibnamefont
  {{Mazloomi M}}}, \bibinfo {author} {\bibfnamefont {S.~S.}\ \bibnamefont
  {Chikatamarla}}, \ and\ \bibinfo {author} {\bibfnamefont {I.~V.}\
  \bibnamefont {Karlin}},\ }\href {\doibase 10.1103/PhysRevLett.114.174502}
  {\bibfield  {journal} {\bibinfo  {journal} {Physical Review Letters}\
  }\textbf {\bibinfo {volume} {114}},\ \bibinfo {pages} {174502} (\bibinfo
  {year} {2015})}\BibitemShut {NoStop}%
\bibitem [{\citenamefont {Kunert}\ and\ \citenamefont
  {Harting}(2007)}]{Kunert2007}%
  \BibitemOpen
  \bibfield  {author} {\bibinfo {author} {\bibfnamefont {C.}~\bibnamefont
  {Kunert}}\ and\ \bibinfo {author} {\bibfnamefont {J.}~\bibnamefont
  {Harting}},\ }\href {\doibase 10.1103/PhysRevLett.99.176001} {\bibfield
  {journal} {\bibinfo  {journal} {Physical Review Letters}\ }\textbf {\bibinfo
  {volume} {99}},\ \bibinfo {pages} {176001} (\bibinfo {year} {2007})},\
  \Eprint {http://arxiv.org/abs/0705.0270} {arXiv:0705.0270} \BibitemShut
  {NoStop}%
\bibitem [{\citenamefont {Hyv{\"{a}}luoma}\ and\ \citenamefont
  {Harting}(2008)}]{Hyvaluoma2008}%
  \BibitemOpen
  \bibfield  {author} {\bibinfo {author} {\bibfnamefont {J.}~\bibnamefont
  {Hyv{\"{a}}luoma}}\ and\ \bibinfo {author} {\bibfnamefont {J.}~\bibnamefont
  {Harting}},\ }\href {\doibase 10.1103/PhysRevLett.100.246001} {\bibfield
  {journal} {\bibinfo  {journal} {Physical Review Letters}\ }\textbf {\bibinfo
  {volume} {100}},\ \bibinfo {pages} {246001} (\bibinfo {year} {2008})},\
  \Eprint {http://arxiv.org/abs/0801.1448} {arXiv:0801.1448} \BibitemShut
  {NoStop}%
\bibitem [{\citenamefont {Atif}\ \emph {et~al.}(2017)\citenamefont {Atif},
  \citenamefont {Kolluru}, \citenamefont {Thantanapally},\ and\ \citenamefont
  {Ansumali}}]{Atif2017}%
  \BibitemOpen
  \bibfield  {author} {\bibinfo {author} {\bibfnamefont {M.}~\bibnamefont
  {Atif}}, \bibinfo {author} {\bibfnamefont {P.~K.}\ \bibnamefont {Kolluru}},
  \bibinfo {author} {\bibfnamefont {C.}~\bibnamefont {Thantanapally}}, \ and\
  \bibinfo {author} {\bibfnamefont {S.}~\bibnamefont {Ansumali}},\ }\href
  {\doibase 10.1103/PhysRevLett.119.240602} {\bibfield  {journal} {\bibinfo
  {journal} {Phys. Rev. Lett.}\ }\textbf {\bibinfo {volume} {119}},\ \bibinfo
  {pages} {240602} (\bibinfo {year} {2017})}\BibitemShut {NoStop}%
\bibitem [{\citenamefont {Dorschner}\ \emph {et~al.}(2016)\citenamefont
  {Dorschner}, \citenamefont {B{\"o}sch}, \citenamefont {Chikatamarla},
  \citenamefont {Boulouchos},\ and\ \citenamefont
  {Karlin}}]{dorschner2016entropic}%
  \BibitemOpen
  \bibfield  {author} {\bibinfo {author} {\bibfnamefont {B.}~\bibnamefont
  {Dorschner}}, \bibinfo {author} {\bibfnamefont {F.}~\bibnamefont
  {B{\"o}sch}}, \bibinfo {author} {\bibfnamefont {S.~S.}\ \bibnamefont
  {Chikatamarla}}, \bibinfo {author} {\bibfnamefont {K.}~\bibnamefont
  {Boulouchos}}, \ and\ \bibinfo {author} {\bibfnamefont {I.~V.}\ \bibnamefont
  {Karlin}},\ }\href@noop {} {\bibfield  {journal} {\bibinfo  {journal}
  {Journal of Fluid Mechanics}\ }\textbf {\bibinfo {volume} {801}},\ \bibinfo
  {pages} {623} (\bibinfo {year} {2016})}\BibitemShut {NoStop}%
\bibitem [{\citenamefont {Pirozzoli}(2011)}]{pirozzoli2011numerical}%
  \BibitemOpen
  \bibfield  {author} {\bibinfo {author} {\bibfnamefont {S.}~\bibnamefont
  {Pirozzoli}},\ }\href@noop {} {\bibfield  {journal} {\bibinfo  {journal}
  {Annual review of fluid mechanics}\ }\textbf {\bibinfo {volume} {43}},\
  \bibinfo {pages} {163} (\bibinfo {year} {2011})}\BibitemShut {NoStop}%
\bibitem [{\citenamefont {Wilde}\ \emph
  {et~al.}(2020{\natexlab{a}})\citenamefont {Wilde}, \citenamefont
  {Kr{\"{a}}mer}, \citenamefont {Reith},\ and\ \citenamefont
  {Foysi}}]{Wilde2020}%
  \BibitemOpen
  \bibfield  {author} {\bibinfo {author} {\bibfnamefont {D.}~\bibnamefont
  {Wilde}}, \bibinfo {author} {\bibfnamefont {A.}~\bibnamefont {Kr{\"{a}}mer}},
  \bibinfo {author} {\bibfnamefont {D.}~\bibnamefont {Reith}}, \ and\ \bibinfo
  {author} {\bibfnamefont {H.}~\bibnamefont {Foysi}},\ }\href {\doibase
  10.1103/PhysRevE.101.053306} {\bibfield  {journal} {\bibinfo  {journal}
  {Physical Review E}\ }\textbf {\bibinfo {volume} {101}},\ \bibinfo {pages}
  {053306} (\bibinfo {year} {2020}{\natexlab{a}})},\ \Eprint
  {http://arxiv.org/abs/1910.13918} {arXiv:1910.13918} \BibitemShut {NoStop}%
\bibitem [{\citenamefont {Feng}\ \emph {et~al.}(2019)\citenamefont {Feng},
  \citenamefont {Boivin}, \citenamefont {Jacob},\ and\ \citenamefont
  {Sagaut}}]{Feng2019}%
  \BibitemOpen
  \bibfield  {author} {\bibinfo {author} {\bibfnamefont {Y.}~\bibnamefont
  {Feng}}, \bibinfo {author} {\bibfnamefont {P.}~\bibnamefont {Boivin}},
  \bibinfo {author} {\bibfnamefont {J.}~\bibnamefont {Jacob}}, \ and\ \bibinfo
  {author} {\bibfnamefont {P.}~\bibnamefont {Sagaut}},\ }\href {\doibase
  10.1016/j.jcp.2019.05.031} {\bibfield  {journal} {\bibinfo  {journal}
  {Journal of Computational Physics}\ }\textbf {\bibinfo {volume} {394}},\
  \bibinfo {pages} {82} (\bibinfo {year} {2019})}\BibitemShut {NoStop}%
\bibitem [{\citenamefont {Frapolli}\ \emph {et~al.}(2015)\citenamefont
  {Frapolli}, \citenamefont {Chikatamarla},\ and\ \citenamefont
  {Karlin}}]{Frapolli2015}%
  \BibitemOpen
  \bibfield  {author} {\bibinfo {author} {\bibfnamefont {N.}~\bibnamefont
  {Frapolli}}, \bibinfo {author} {\bibfnamefont {S.~S.}\ \bibnamefont
  {Chikatamarla}}, \ and\ \bibinfo {author} {\bibfnamefont {I.~V.}\
  \bibnamefont {Karlin}},\ }\href {\doibase 10.1103/PhysRevE.92.061301}
  {\bibfield  {journal} {\bibinfo  {journal} {Physical Review E - Statistical,
  Nonlinear, and Soft Matter Physics}\ }\textbf {\bibinfo {volume} {92}},\
  \bibinfo {pages} {061301} (\bibinfo {year} {2015})}\BibitemShut {NoStop}%
\bibitem [{\citenamefont {Prasianakis}\ and\ \citenamefont
  {Karlin}(2008)}]{Prasianakis2008}%
  \BibitemOpen
  \bibfield  {author} {\bibinfo {author} {\bibfnamefont {N.~I.}\ \bibnamefont
  {Prasianakis}}\ and\ \bibinfo {author} {\bibfnamefont {I.~V.}\ \bibnamefont
  {Karlin}},\ }\href {\doibase 10.1103/PhysRevE.78.016704} {\bibfield
  {journal} {\bibinfo  {journal} {Physical Review E - Statistical, Nonlinear,
  and Soft Matter Physics}\ }\textbf {\bibinfo {volume} {78}},\ \bibinfo
  {pages} {016704} (\bibinfo {year} {2008})}\BibitemShut {NoStop}%
\bibitem [{\citenamefont {Frapolli}\ \emph
  {et~al.}(2016{\natexlab{a}})\citenamefont {Frapolli}, \citenamefont
  {Chikatamarla},\ and\ \citenamefont {Karlin}}]{Frapolli2016}%
  \BibitemOpen
  \bibfield  {author} {\bibinfo {author} {\bibfnamefont {N.}~\bibnamefont
  {Frapolli}}, \bibinfo {author} {\bibfnamefont {S.~S.}\ \bibnamefont
  {Chikatamarla}}, \ and\ \bibinfo {author} {\bibfnamefont {I.~V.}\
  \bibnamefont {Karlin}},\ }\href@noop {} {\bibfield  {journal} {\bibinfo
  {journal} {Physical review letters}\ }\textbf {\bibinfo {volume} {117}},\
  \bibinfo {pages} {10604} (\bibinfo {year} {2016}{\natexlab{a}})}\BibitemShut
  {NoStop}%
\bibitem [{\citenamefont {Frapolli}\ \emph
  {et~al.}(2016{\natexlab{b}})\citenamefont {Frapolli}, \citenamefont
  {Chikatamarla},\ and\ \citenamefont {Karlin}}]{frapolli2016entropic}%
  \BibitemOpen
  \bibfield  {author} {\bibinfo {author} {\bibfnamefont {N.}~\bibnamefont
  {Frapolli}}, \bibinfo {author} {\bibfnamefont {S.~S.}\ \bibnamefont
  {Chikatamarla}}, \ and\ \bibinfo {author} {\bibfnamefont {I.~V.}\
  \bibnamefont {Karlin}},\ }\href@noop {} {\bibfield  {journal} {\bibinfo
  {journal} {Physical Review E}\ }\textbf {\bibinfo {volume} {93}},\ \bibinfo
  {pages} {63302} (\bibinfo {year} {2016}{\natexlab{b}})}\BibitemShut {NoStop}%
\bibitem [{\citenamefont {Chikatamarla}\ and\ \citenamefont
  {Karlin}(2006)}]{shyam2006prl}%
  \BibitemOpen
  \bibfield  {author} {\bibinfo {author} {\bibfnamefont {S.~S.}\ \bibnamefont
  {Chikatamarla}}\ and\ \bibinfo {author} {\bibfnamefont {I.~V.}\ \bibnamefont
  {Karlin}},\ }\href@noop {} {\bibfield  {journal} {\bibinfo  {journal}
  {Physical review letters}\ }\textbf {\bibinfo {volume} {97}},\ \bibinfo
  {pages} {190601} (\bibinfo {year} {2006})}\BibitemShut {NoStop}%
\bibitem [{\citenamefont {Dorschner}\ \emph {et~al.}(2018)\citenamefont
  {Dorschner}, \citenamefont {B{\"{o}}sch},\ and\ \citenamefont
  {Karlin}}]{pond}%
  \BibitemOpen
  \bibfield  {author} {\bibinfo {author} {\bibfnamefont {B.}~\bibnamefont
  {Dorschner}}, \bibinfo {author} {\bibfnamefont {F.}~\bibnamefont
  {B{\"{o}}sch}}, \ and\ \bibinfo {author} {\bibfnamefont {I.~V.}\ \bibnamefont
  {Karlin}},\ }\href {\doibase 10.1103/PhysRevLett.121.130602} {\bibfield
  {journal} {\bibinfo  {journal} {Physical Review Letters}\ }\textbf {\bibinfo
  {volume} {121}},\ \bibinfo {pages} {130602} (\bibinfo {year} {2018})},\
  \Eprint {http://arxiv.org/abs/1806.05089} {arXiv:1806.05089} \BibitemShut
  {NoStop}%
\bibitem [{\citenamefont {He}\ and\ \citenamefont {Luo}(1997)}]{He-Lou-1997}%
  \BibitemOpen
  \bibfield  {author} {\bibinfo {author} {\bibfnamefont {X.}~\bibnamefont
  {He}}\ and\ \bibinfo {author} {\bibfnamefont {L.-S.}\ \bibnamefont {Luo}},\
  }\href {\doibase 10.1103/PhysRevE.56.6811} {\bibfield  {journal} {\bibinfo
  {journal} {Phys. Rev. E}\ }\textbf {\bibinfo {volume} {56}},\ \bibinfo
  {pages} {6811} (\bibinfo {year} {1997})}\BibitemShut {NoStop}%
\bibitem [{\citenamefont {Ansumali}\ and\ \citenamefont
  {Karlin}(2005)}]{ansumali2005prl}%
  \BibitemOpen
  \bibfield  {author} {\bibinfo {author} {\bibfnamefont {S.}~\bibnamefont
  {Ansumali}}\ and\ \bibinfo {author} {\bibfnamefont {I.~V.}\ \bibnamefont
  {Karlin}},\ }\href {\doibase 10.1103/PhysRevLett.95.260605} {\bibfield
  {journal} {\bibinfo  {journal} {Phys. Rev. Lett.}\ }\textbf {\bibinfo
  {volume} {95}},\ \bibinfo {pages} {260605} (\bibinfo {year}
  {2005})}\BibitemShut {NoStop}%
\bibitem [{\citenamefont {Koumoutsakos}(1997)}]{koumoutsakos1997inviscid}%
  \BibitemOpen
  \bibfield  {author} {\bibinfo {author} {\bibfnamefont {P.}~\bibnamefont
  {Koumoutsakos}},\ }\href@noop {} {\bibfield  {journal} {\bibinfo  {journal}
  {Journal of Computational Physics}\ }\textbf {\bibinfo {volume} {138}},\
  \bibinfo {pages} {821} (\bibinfo {year} {1997})}\BibitemShut {NoStop}%
\bibitem [{\citenamefont {van Rees}(2014)}]{rees20143d}%
  \BibitemOpen
  \bibfield  {author} {\bibinfo {author} {\bibfnamefont {W.~M.}\ \bibnamefont
  {van Rees}},\ }\emph {\bibinfo {title} {{3D simulations of vortex dynamics
  and biolocomotion}}},\ \href@noop {} {Ph.D. thesis},\ \bibinfo  {school} {ETH
  Zurich} (\bibinfo {year} {2014})\BibitemShut {NoStop}%
\bibitem [{\citenamefont {Kr{\"{a}}mer}\ \emph {et~al.}(2017)\citenamefont
  {Kr{\"{a}}mer}, \citenamefont {K{\"{u}}llmer}, \citenamefont {Reith},
  \citenamefont {Joppich},\ and\ \citenamefont {Foysi}}]{kramer2017semi}%
  \BibitemOpen
  \bibfield  {author} {\bibinfo {author} {\bibfnamefont {A.}~\bibnamefont
  {Kr{\"{a}}mer}}, \bibinfo {author} {\bibfnamefont {K.}~\bibnamefont
  {K{\"{u}}llmer}}, \bibinfo {author} {\bibfnamefont {D.}~\bibnamefont
  {Reith}}, \bibinfo {author} {\bibfnamefont {W.}~\bibnamefont {Joppich}}, \
  and\ \bibinfo {author} {\bibfnamefont {H.}~\bibnamefont {Foysi}},\
  }\href@noop {} {\bibfield  {journal} {\bibinfo  {journal} {Physical Review
  E}\ }\textbf {\bibinfo {volume} {95}},\ \bibinfo {pages} {23305} (\bibinfo
  {year} {2017})}\BibitemShut {NoStop}%
\bibitem [{\citenamefont {Wilde}\ \emph
  {et~al.}(2020{\natexlab{b}})\citenamefont {Wilde}, \citenamefont
  {Kr{\"{a}}mer}, \citenamefont {Reith},\ and\ \citenamefont
  {Foysi}}]{wilde2020semi}%
  \BibitemOpen
  \bibfield  {author} {\bibinfo {author} {\bibfnamefont {D.}~\bibnamefont
  {Wilde}}, \bibinfo {author} {\bibfnamefont {A.}~\bibnamefont {Kr{\"{a}}mer}},
  \bibinfo {author} {\bibfnamefont {D.}~\bibnamefont {Reith}}, \ and\ \bibinfo
  {author} {\bibfnamefont {H.}~\bibnamefont {Foysi}},\ }\href@noop {}
  {\bibfield  {journal} {\bibinfo  {journal} {Physical Review E}\ }\textbf
  {\bibinfo {volume} {101}},\ \bibinfo {pages} {53306} (\bibinfo {year}
  {2020}{\natexlab{b}})}\BibitemShut {NoStop}%
\bibitem [{\citenamefont {Kr{\"a}mer}\ \emph {et~al.}(2020)\citenamefont
  {Kr{\"a}mer}, \citenamefont {Wilde}, \citenamefont {K{\"u}llmer},
  \citenamefont {Reith}, \citenamefont {Foysi},\ and\ \citenamefont
  {Joppich}}]{kramer2020lattice}%
  \BibitemOpen
  \bibfield  {author} {\bibinfo {author} {\bibfnamefont {A.}~\bibnamefont
  {Kr{\"a}mer}}, \bibinfo {author} {\bibfnamefont {D.}~\bibnamefont {Wilde}},
  \bibinfo {author} {\bibfnamefont {K.}~\bibnamefont {K{\"u}llmer}}, \bibinfo
  {author} {\bibfnamefont {D.}~\bibnamefont {Reith}}, \bibinfo {author}
  {\bibfnamefont {H.}~\bibnamefont {Foysi}}, \ and\ \bibinfo {author}
  {\bibfnamefont {W.}~\bibnamefont {Joppich}},\ }\href@noop {} {\bibfield
  {journal} {\bibinfo  {journal} {Computers \& Mathematics with Applications}\
  }\textbf {\bibinfo {volume} {79}},\ \bibinfo {pages} {34} (\bibinfo {year}
  {2020})}\BibitemShut {NoStop}%
\bibitem [{\citenamefont {INOUE}\ and\ \citenamefont
  {HATTORI}(1999)}]{inoue_hattori_1999}%
  \BibitemOpen
  \bibfield  {author} {\bibinfo {author} {\bibfnamefont {O.}~\bibnamefont
  {INOUE}}\ and\ \bibinfo {author} {\bibfnamefont {Y.}~\bibnamefont
  {HATTORI}},\ }\href {\doibase 10.1017/S0022112098003565} {\bibfield
  {journal} {\bibinfo  {journal} {Journal of Fluid Mechanics}\ }\textbf
  {\bibinfo {volume} {380}},\ \bibinfo {pages} {81–116} (\bibinfo {year}
  {1999})}\BibitemShut {NoStop}%
\end{thebibliography}

\end{document}